\documentclass{iopart}
\usepackage{graphicx}
\begin{document}
\review[Order Parameters and Phase Diagrams of Multiferroics]
{Order Parameters and Phase Diagrams of Multiferroics}
\author{A. B. Harris$^1$, Amnon Aharony$^2$, and Ora Entin-Wohlman$^2$}
\address{$ˆ1$ Department of Physics and Astronomy, University
of Pennsylvania, Philadelphia, PA 19104, USA}
\address{$ˆ2$ Department of Physics, Ben Gurion University,
Beer Sheva 84105 ISRAEL$^*$} \ead{aaharony@bgu.ac.il}
\begin{abstract}
The symmetry properties, order parameters, and magnetoelectric phase
diagrams of multiferroics are discussed. After brief reviews of
Ni$_3$V$_2$O$_8$, TbMnO$_3$, and RbFe(MoO$_4$)$_2$, we present a
detailed analysis of RMn$_2$O$_5$ (with R=Y, Ho, Dy, Er, Tb, Tm).
\end{abstract}
\submitto{\JPCM}
\maketitle
\def\rhov{{\mbox{\boldmath{$\rho$}}}}
\def\tauv{{\mbox{\boldmath{$\tau$}}}}
\def\Lambdav{{\mbox{\boldmath{$\Lambda$}}}}
\def\sigmav{{\mbox{\boldmath{$\sigma$}}}}
\def\xiv{{\mbox{\boldmath{$\xi$}}}}
\def\chiv{{\mbox{\boldmath{$\chi$}}}}
\def\oh{{\scriptsize 1 \over \scriptsize 2}}
\def\ot{{\scriptsize 1 \over \scriptsize 3}}
\def\of{{\scriptsize 1 \over \scriptsize 4}}
\def\tf{{\scriptsize 3 \over \scriptsize 4}}

\section{INTRODUCTION}

Here we review recent and new developments which elucidate the
symmetry and the description of the magnetic and dielectric states
of multiferroics using order parameters.  After some examples
where the magnetoelectric (ME) behavior is relatively simple, we
discuss a proposed generic phase diagram for the so-called ``125"
systems, RMn$_2$O$_5$, where R is a rare earth. The most important
consequence of the phenomenological theories we develop is to
provide a general framework for understanding the magnetic and
dielectric properties of these materials and how these properties
combine to produce the interesting ME phenomena.

Briefly, this article is organized as follows.  In Sec. 2 we
discuss the characterization of the magnetic structure obtained
from symmetry arguments.  Here we discuss briefly a simplified
version of the group theoretical approach (known as representation
theory) which is supplemented by less well-known arguments
involving the use of inversion symmetry. As examples we consider
Ni$_3$V$_2$O$_8$ (NVO), TbMnO$_3$, and RbFe(MoO$_4$)$_2$ (RFMO)
and discuss the introduction of order parameters (OP's) to
characterize the magnetic symmetry. We then give a brief review of
how symmetry restricts the form of the ME interaction when it is
written in terms of both magnetic and dielectric OP's. In Sec. 3
we give a detailed discussion of how these concepts enable us to
construct a generic phase diagram for the 125 family of
multiferroics, which does not rely on a knowledge of the details
of the microscopic interactions. Section 4  contains an application of the theory
of critical phenomena to the 125's, and the
paper is briefly summarized in section 5.

\section{SYMMETRY AND MAGNETIC STRUCTURE}

Here we give a simplified review of the role of symmetry in
determining the structure of the magnetically ordered phase which
develops {\it at a continuous phase transition}.  This subject is
of ancient vintage, being discussed about 60 years ago by Landau
(see \cite{LDL}). However, some reviews which discuss the analysis
of diffraction data\cite{EFB,RM} overlook the importance of
inversion symmetry in reducing the number of parameters needed to
describe the ordered magnetic structures. For multiferroics this
was first corrected quite recently by Lawes {\it et
al.}\cite{GL05}, by Kenzelmann {\it et al.}\cite{MK05} and in more
detail by Harris\cite{ABH07a}, which we follow here. Formal
treatments appeared some time ago \cite{newrev}. Recent papers include
Schweizer {\it et al.}\cite{JS07} and Radaelli and
Chapon\cite{PR07}.

\begin{figure}[ht]
\begin{center}
\caption{(Color online).  (a) The six Ni sites in the unit cell of
NVO. Circles represent ``spine" sites and squares represent the
``cross-tie" sites. The axis of the two-fold rotation about the
$a$-axis is shown. The glide plane is indicated by the mirror
plane at $z= \tf$ and the arrow above $m_c$ indicates that a
translation of $b/2$ in the $b$-direction is involved. (b) The
four Mn sites (small circles) and four Tb sites (large circles) in
the unit cell of TbMnO$_3$. The glide $m_a$ is indicated by the
mirror plane at $x=\tf$ followed by a translation of $b/2$ along
the ${\bf b}$ direction. The planes at $z=1/4$ and $z=3/4$ are
mirror planes. (c) RFMO, where large balls are Fe spins 5/2 on a
stacked triangular lattice, small balls are oxygens, Mo ions are
inside the oxygen tetrahedra, and the Rb ions are not shown.}
\label{STRUCT} \vspace{0.1 in}

\includegraphics[width=3.8cm]{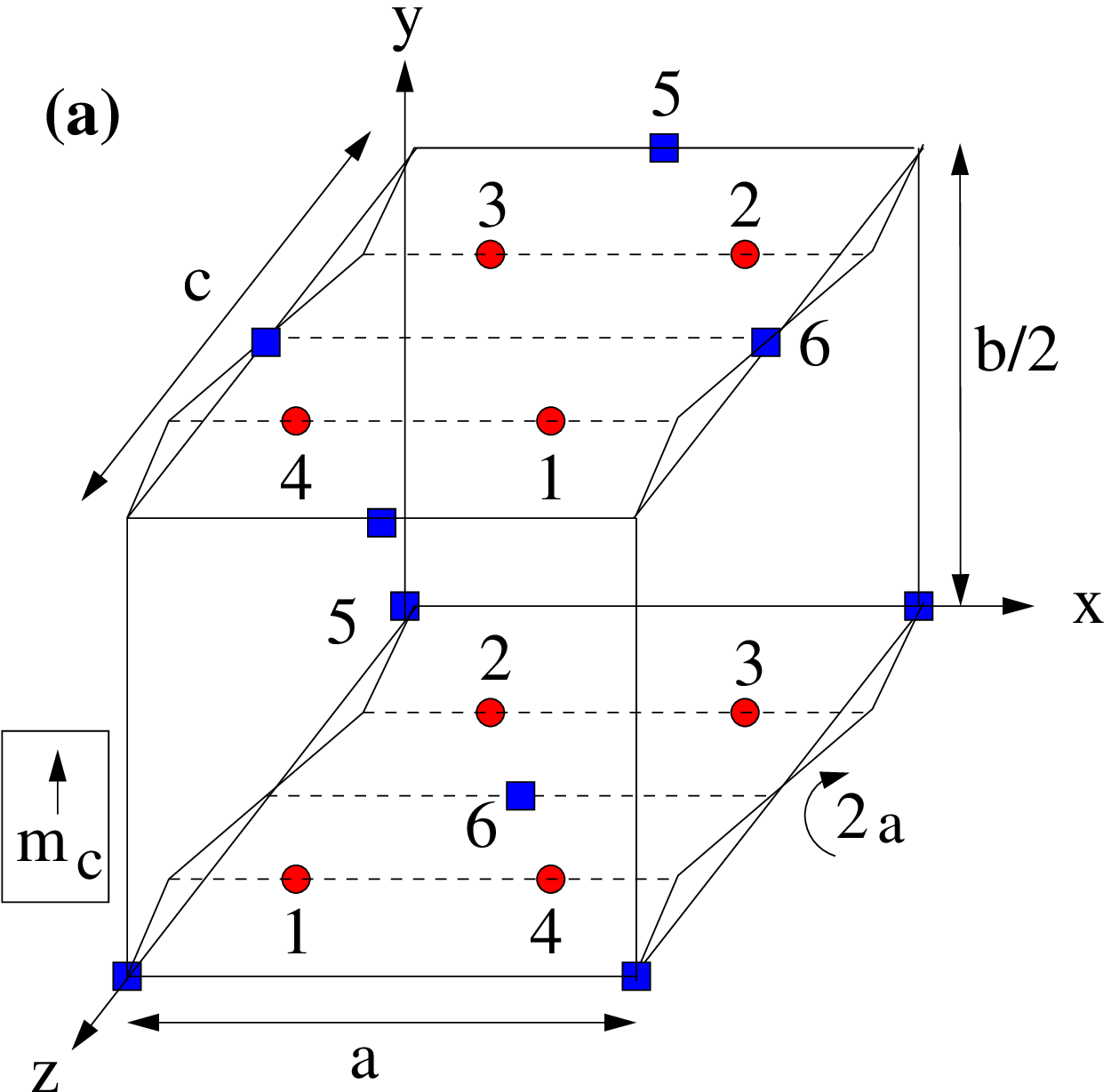}
\includegraphics[width=4.3cm]{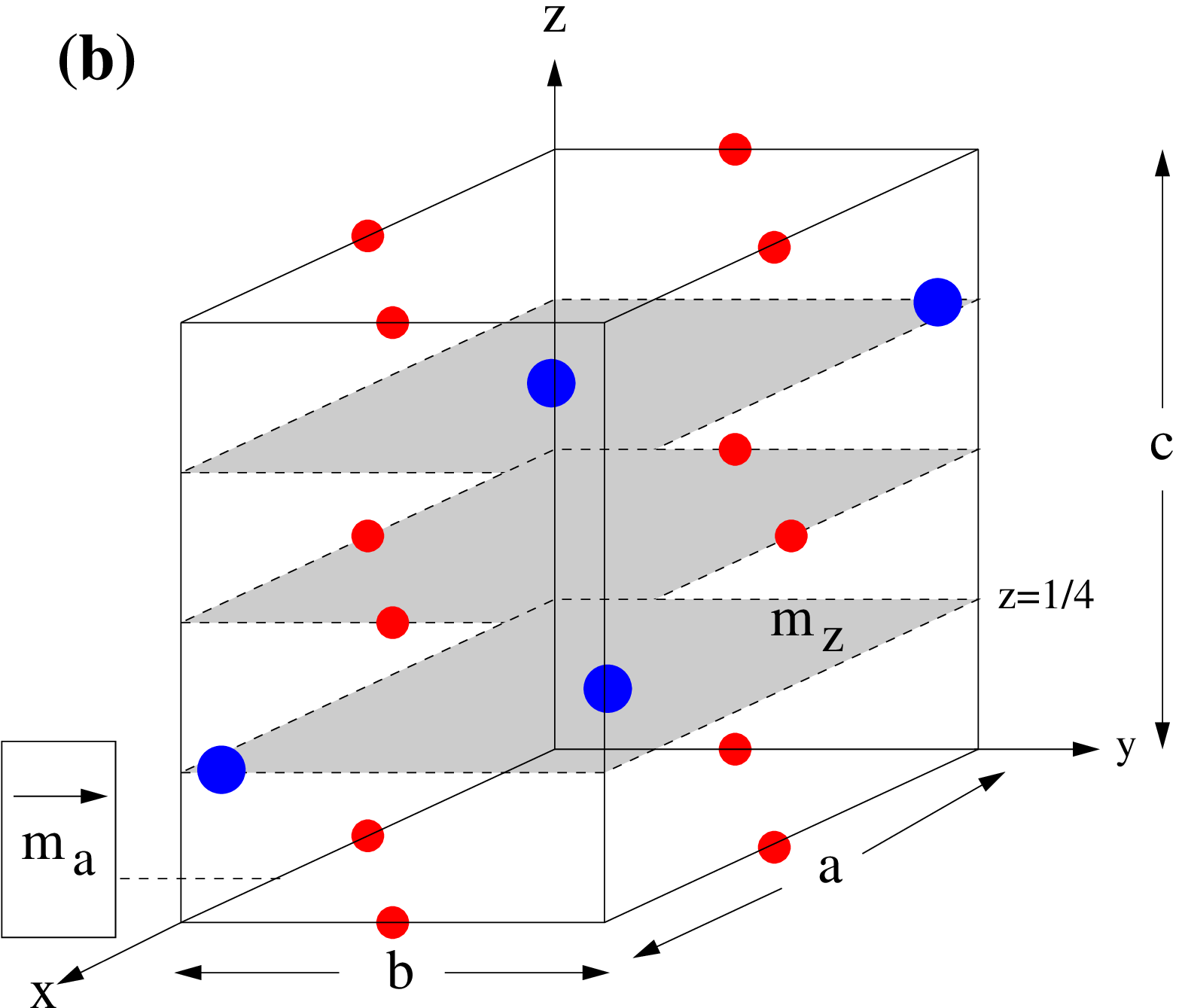}
\hspace{0.1 in}
\includegraphics[width=3.4cm]{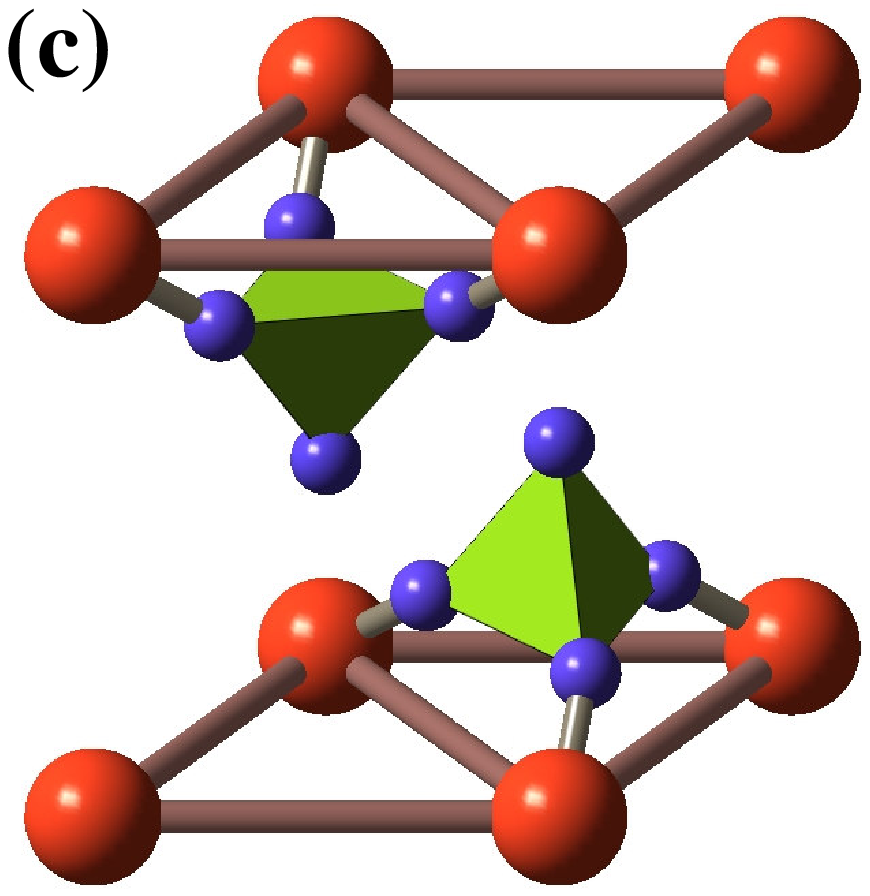}
\end{center}
\end{figure}

We start by assuming that the paramagnetic phase is characterized by a
primitive unit cell with $n_\tau$ magnetic sites. The Landau expansion of $F_2$,
the magnetic free energy at quadratic order in the spin components, is
\begin{eqnarray}
F_2 &=& \sum_{\bf q} \sum_{\tau, \tau'=1}^{n_\tau}
\sum_{\alpha , \beta} [\chiv^{-1}({\bf q})]_{\tau, \alpha ; \tau', \beta}
S_{\alpha}({\bf q},\tau)^* S_{\beta}({\bf q},\tau')  \ ,
\label{F2EQ} \end{eqnarray}
where $\chiv$ is the wave-vector dependent susceptibility matrix and
\begin{eqnarray}
S_\alpha ({\bf R}, \tau) &=& \sum_{\bf q}S_\alpha ({\bf q}, \tau)
e^{i {\bf q} \cdot {\bf R}} \ ,
\end{eqnarray}
where $S_{\alpha}({\bf R},\tau)$ is the $\alpha$-component of spin
of the $\tau$th magnetic site in the unit cell at ${\bf R}$ and
$S_\alpha (-{\bf q},\tau)=S_\alpha({\bf q},\tau)^*$. For each
value of the wave vector the inverse susceptibility has $3n_\tau$
eigenvalues (which may or may not be distinct from one another).
At high temperature $T$ all these eigenvalues are positive and the
paramagnetic state is thermodynamically stable. As
$T$ is reduced through a critical value, $T_c$, one eigenvalue,
$\lambda_c({\bf q}_c)$, at some wave vector ${\bf q}_c$ (and wave
vectors equivalent to it by symmetry which comprise the {\it star}
of ${\bf q}_c$) approaches zero, signaling an instability of the
paramagnetic phase to the formation of long-range order at the
critical wave vector ${\bf q}_c$ associated with this critical
eigenvalue.  The actual value of ${\bf q}_c$ is determined by the
microscopic interactions. Since these interactions are not well
known, we regard the wave vector as an experimentally determined
parameter.  The degeneracy of this critical eigenvalue $\lambda_c$
is $n_q{\cal N}$, where $n_q$ is the number of wave vectors in the
star of ${\bf q}$ and ${\cal N}$ is the dimensionality of the
irreducible representation (irrep) of the symmetry group (the
so-called "little group") which leaves the wave vector invariant.
(For ferromagnetic Ising, $x$-$y$, and Heisenberg models
${\cal N}$ assumes the values 1, 2, and 3, respectively.) To avoid
technicalities, in this section we consider the simplest
case, ${\cal N}=1$. This case is simple because then we can
use the familiar principle that the eigenvectors of a matrix (here
the inverse susceptibility) are also simultaneously eigenvectors
of operators (here the symmetry operations ${\cal O}_i$ of the
space group which leave the selected wave vector invariant) which
commute with each other and with the matrix. In this way we avoid
using the full apparatus of group theory and the reader need not
know anything at all about ``irreps".  We now illustrate this
idea and show how inversion symmetry introduces further
simplifications for three recently studied multiferroic magnetic
materials, whose lattice structures are shown in figure
\ref{STRUCT} and whose positions (except for RFMO where the Fe
ions form a Bravais lattice) are given in table \ref{SITES1}.
{\footnote{We interchangeably denote the ${\bf a}$, ${\bf b}$, and
${\bf c}$ axes as $x$, $y$, and $z$, respectively.}}

\begin{table}
\fl
\caption{\label{SITES1}General positions\cite{ITC} (given as fractions
of lattice constants) within the primitive unit cell for NVO (space
group Cmca) and TbMnO$_3$ (space group Pbnm). Here $r+\equiv r+1/2$,
$2_\alpha$ is a two-fold rotation (or screw) axis, and $m_\alpha$ is
a mirror (or glide).}
\vspace{0.2 in}

\begin{tabular} {||c | c ||}
\hline \hline \multicolumn {2} {||c||} {Ni$_3$V$_2$O$_8$} \\ \hline
${\rm E}{\bf r} =(x,y,z)$ & $2_c{\bf r} =(\overline x,
\overline y +, z +)$ \\ $2_a {\bf r} = (x, \overline y , \overline z)$
& $2_b {\bf r} = (\overline x, y+, \overline z+)$ \\ ${\cal I} {\bf r}
=(\overline x,\overline y,\overline z)$ &$m_c{\bf r} =(x, y+, \overline z+)$ \\
$m_a{\bf r} = (\overline x, y , z)$ & $m_b{\bf r} = (x, \overline y+, z+)$ \\
\hline \hline \end{tabular}
\begin{tabular} {|| c | c ||}
\hline \hline \multicolumn {2} {||c||} {TbMnO$_3$} \\ \hline $E {\bf r} =(x,y,z) \ \ $ & $2_a {\bf r} =(x+, \overline y+, \overline z )\ \ $ \\
$2_c{\bf r} =(\overline x,\overline y , z+)\ \ $ & $2_b{\bf r} =(\overline x+, y+, \overline z+)\ \ $ \\ ${\cal I}{\bf r}=(\overline x,\overline
y,\overline z)\ \ $ & $m_a{\bf r}=(\overline x+, y+, z )\ \ $ \\ $m_c{\bf r}=(x, y ,\overline z+)\ \ $ & $m_b{\bf r}=(x+, \overline y+, z+)\ \ $ \\
\hline \hline
\end{tabular}
\end{table}

\subsection{NVO}

For NVO the incommensurate (IC) wave vector for magnetic ordering
is\cite{GL04,MK06} ${\bf q}\cong 0.28 (2 \pi /a) \hat {\bf a}$. Thus the
space group operations ${\cal O}_i$ which leave the wave vector
invariant are generated by $2_a$, a two-fold rotation about
the $a$-axis and passes through the
origin and $m_c$, a glide operation which takes $c$ into $-c$
followed by a translation through $(b/2) \hat {\bf b}$. Thus the
critical eigenvector (which is the spatial Fourier transform of
the spin distribution) must not only be an eigenvector of  the
inverse susceptibility matrix, but it must also simultaneously be
an eigenvector of both $2_a$ and $m_c$. Since $[2_a]^2=1$, the
eigenvalues of $2_a$ must be $\lambda (2_a) \equiv \lambda=\pm 1$.
Since $[m_c]^2$ is a translation along the $b$ axis, the eigenvalues
of $m_c$ must be $\lambda(m_c) \equiv \lambda'=\pm \exp(ib q_b/2)=\pm 1$.
Thus, if we assume continuous transitions, there can only be four
distinct symmetries of ordered phases, corresponding to
independently selecting the eigenvalues of $2_a$ and $m_c$.  The
corresponding eigenvectors must be of the form
\begin{eqnarray}
\fl \hspace{0.5 in}
S(q,1)&=&(\alpha_1 , \alpha_2, \alpha_3)\xi \ , \ \ \
S(q,2)=\lambda(\alpha_1, - \alpha_2 , - \alpha_3 )\xi \ , \nonumber \\
\fl \hspace{0.5 in}
S(q,3)&=&\lambda \lambda'(-\alpha_1 , \alpha_2, - \alpha_3 )\xi^3 \ ,
\ \ \ S(q,4)=\lambda'(-\alpha_1 , - \alpha_2, + \alpha_3 ) /\xi^3 \ ,
\nonumber \\
\fl \hspace{0.5 in}
 S(q,5)&=&( [1+\lambda]\alpha_4 , [1-\lambda]\alpha_5, [1-\lambda]
\alpha_6) \ , \nonumber \\
\fl \hspace{0.5 in} S(q,6) &=& - \lambda' ([1+\lambda]\alpha_4,
[1-\lambda] \alpha_5, [1-\lambda]\alpha_6)\xi^2 \ , \label{IRREP}
\end{eqnarray}
where $\xi=\exp(iq_xa/4)$ and the $\alpha_n$ assume arbitrary
complex values. To check this note that under $2_a$ sublattices
\#1 and \#2 are interchanged as are \#3 and \#4, whereas under
$m_c$ sublattices \#1 and \#4 are interchanged as are \#2 and \#3.
Note that $2_a$ changes the signs of the $b$ and $c$-components of
spin, while $m_c$ changes the signs of the $a$ and $b$ components
of spin since spin is a pseudo-vector. This type of analysis,
known as representation theory, is well-known and widely used.
However, less well-known and often overlooked (as documented in
\cite{ABH07a}) is the fact that in these multiferroic systems the
free energy must be invariant under the inversion symmetry ${\cal
I}$ possessed by the lattice \cite{newrev}. One can then
show\cite{GL05,ABH06b,MK06,ABH07a} that this symmetry fixes the
phases of the $\alpha_n$: for $\lambda=\lambda'=1$, apart from an
overall complex phase factor, $\alpha_1$ and $\alpha_3$ must be
pure imaginary and $\alpha_2$ and $\alpha_4$ must be pure real.
For other irreps [i. e. for the three other choices of the
eigenvalues $\lambda ({\cal O}_i)$] one has analogous results. If
(\ref{F2EQ}) is generalized to include terms of fourth order in
the spin variables, then a mean-field analysis for $T$ near $T_c$
shows that the overall amplitude of the spin wave function varies
[proportionally to $(T_c-T)^{1/2}$], but the ratios among the
$\alpha_n$'s are nearly temperature independent. Therefore we
replace $\alpha_n$ by $\sigma ({\bf q}) \alpha_n$ and require the
normalization $\sum | \alpha_n|^2=1$.  Thus the temperature
dependence is incorporated in the order parameter $\sigma$. If we
require that $\alpha_4$, say, be real, then the freedom to fix the
overall phase is taken into account by allowing the order
parameter to be complex, as one would expect, since the origin of
the IC ordering is not fixed, at least within $F_2$.  It should be
noted that the order parameter inherits the symmetry of the spin
functions, so that
\begin{eqnarray}
2_a \sigma = \lambda (2_a) \sigma =\lambda\sigma\ , \ \ \ m_c
\sigma = \lambda (m_c) \sigma =\lambda'\sigma\ , \ \ \ {\cal I}
\sigma = \sigma^* \ . \label{NVOEQ} \end{eqnarray}

In the analysis of diffraction experiments one tries to fit the structure
assuming in turn each of the four symmetries. In so doing one has not
$3n_\tau=18$ complex-valued fitting parameters, but rather the 4 or 5
$\alpha$'s of (\ref{IRREP}) (depending on which symmetry one is considering).
However, the use of inversion further reduces the number of fitting
parameters by half since their phases are fixed \cite{MK06}.

\begin{figure}[ht]
\begin{center}
\caption{Dielectric and magnetic phase diagrams of NVO (left, from
\cite{GL04,GL05,MK06}), TbMnO$_3$ (center, from \cite{TK05,MK05}),
and RFMO (right, from \cite{LES03,MK07}). In the dielectric phase
diagram the direction of the spontaneous polarization (if any) is
indicated. For NVO $T_C\approx 4$K, $T_<\approx 6$K, and $T_>
\approx 9$K.  In the magnetic phase diagrams P denotes
paramagnetic, HTI denotes a dominantly collinear IC phase with a
single OP,  LTI is a dominantly elliptically polarized phase with two
OP's, and IC-TRI denotes the IC stacking of triangular lattice
antiferromagnets.} \label{PDF} \vspace{0.2 in}

\includegraphics[width=4cm]{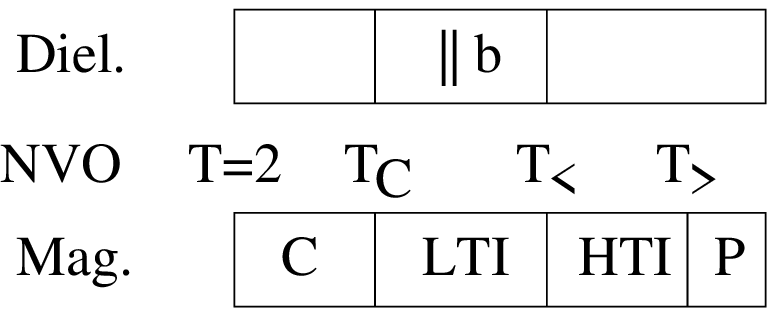}
\includegraphics[width=4cm]{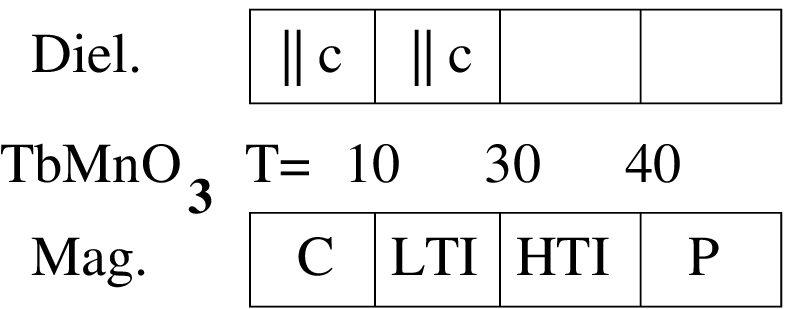}
\includegraphics[width=4cm]{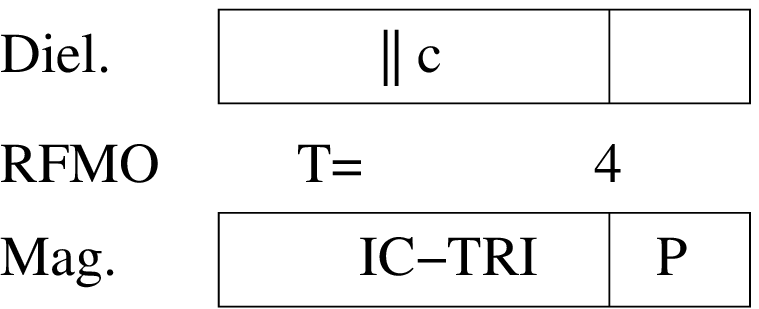}
\end{center}
\end{figure}

The left phase diagram in figure \ref{PDF} shows the experimentally observed sequence of
magnetic phases of NVO. As $T$ is lowered, the first
ordered phase to appear (at $T_> \approx 9$K) is the HTI phase,
which has a single eigenvector associated with predominantly
collinear sinusoidally modulated order.  Analysis of experimental
data indicates that the eigenvalues of this phase are\cite{MK06}
$\lambda(2_x)=-1$ and $\lambda(m_c)=+1$.  At a lower temperature
($T_< \approx 6$K)  the LTI phase appears with an {\it additional}
order parameter associated with dominantly transverse spin order and with
$\lambda(2_x)=+1$ and $\lambda(m_c)=+1$.  The magnetic free energy
which describes the development of these two successive ordering
transitions is of the form\cite{MK06,ABH07a,ABH08a}
\begin{eqnarray}
F_M &=& a(T-T_>) |\sigma_{\rm HTI}|^2 + b(T-T_<) |\sigma_{\rm LTI}|^2
+ {\cal O} (\sigma^4)\ ,
\label{FREEMAG} \end{eqnarray}
where $a$ and $b$ are constants and $T_>$ and $T_<$ are the respective
temperatures (when nonquadratic terms are ignored) at which
$\sigma_{\rm HTI}$ and $\sigma_{\rm LTI}$ become nonzero.  The unwritten
terms in $F_M$, which are quartic in $\sigma$, favor fixed spin length.
Thus $\sigma_{\rm LTI}$ is out of phase relative to $\sigma_{\rm HTI}$
and the spins thereby order in a spiral structure\cite{MK06,TK61,TN67}.

\subsection{TbMnO$_3$}

The case of TbMnO$_3$ is almost identical to that for NVO.
Here the IC wave vector is of the form $(0,q,0)$\cite{RK04,MK05},
so that the symmetry operations which leave it invariant are
generated by the glide $m_a$ and the mirror $m_c$. The eigenvalues
of $m_c$ ($m_a$) are $\pm 1$ ($\pm \Lambda$),
where $\Lambda=\exp(i \pi q)$. For $\lambda(m_c)=1$ and
$\lambda(m_a)=\Lambda$ one has
\begin{eqnarray}
&& S(q,1)=\alpha_1 \hat i - \alpha_2 \hat j - \alpha_3 \hat k \ ,
\ \ \ S(q,2)=\alpha_1 \hat i + \alpha_2 \hat j + \alpha_3 \hat k \ ,
\nonumber \\ &&
S(q,3)=-\alpha_1 \hat i + \alpha_2 \hat j - \alpha_3 \hat k \ , \ \ \
S(q,4)=-\alpha_1 \hat i - \alpha_2 \hat j + \alpha_3 \hat k \ ,
\nonumber \\ &&
S(q,5)=\alpha_4 \hat k \ , \ \ \ S(q,6)=-\alpha_5 \hat k \ , \nonumber\\
&&S(q,7)=\alpha_5 \hat k \ , \ \ \ S(q,8)=-\alpha_4 \hat k \ ,
\end{eqnarray}
where the $\alpha_n$ are arbitrary complex numbers. As for NVO one
can now require that $F_2$ be invariant under ${\cal I}$. In this
case the result is that apart from an overall phase factor,
$\alpha_n$ for $n=1,2,3$ are real, $\alpha_4$ is an arbitrary
complex number, and
$\alpha_5=\alpha_4^*$\cite{MK05,ABH06b,ABH07a}. When inversion
symmetry relates sites within the same Wyckoff orbit of the
operators of the group of the wave vector (as it does for NVO),
the complex phases get fixed, whereas when inversion relates sites
in {\it different} Wyckoff orbits (as for the Tb sites here), the
previously independent amplitudes of the two orbits are now
related. Again, we replace $\alpha_n$ by $\sigma ({\bf q})
\alpha_n$, so that the temperature dependence of the spin function
is essentially contained in the order parameter $\sigma({\bf q})$
and the complex phase of the IC ordering is incorporated in the
arbitrary complex phase of $\sigma({\bf q})$ which transforms as
\begin{eqnarray}
m_a \sigma ({\bf q}) = \lambda (m_a) \sigma ({\bf q}) \ , \ \
m_c \sigma ({\bf q}) = \lambda (m_c) \sigma ({\bf q}) \ , \ \
{\cal I} \sigma ({\bf q}) = \sigma ({\bf q})^* \ .
\label{TMOEQ} \end{eqnarray}

The center phase diagram in figure \ref{PDF} shows the sequence of
magnetic phases of TbMnO$_3$. As the temperature is lowered
(through $T_>=40$K) the first ordered phase to appear is the HTI
phase in which the single eigenvector associated with
predominantly collinear order appears with $\lambda(m_c)=- \exp(i
\pi q) \equiv -\Lambda$ and $\lambda(m_a)=1$.  At a lower
temperature ($T_<\approx 30$K)  the LTI phase appears with an
additional order parameter associated with transverse spin order
and with $\lambda(m_c)=-\Lambda$ and $\lambda(m_a)=-1$. The
phenomenology of the magnetic phase diagram of TbMnO$_3$ is very
similar to that of NVO.

\subsection{RFMO}\label{RFMO}

The magnetic Fe ions in RFMO form triangular lattice planes which
are stacked directly over one another, as shown in figure
\ref{STRUCT}(c)\cite{LES03}. Below $T=180$K but above the magnetic
ordering temperature (at $T_c=4$K) the lattice has P$\overline 3$
symmetry\cite{GG04}, so that the only symmetry operation (apart
from ${\cal I}$) is a three-fold rotation ${\cal R}$ about the
$c$-axis, which is perpendicular to the triangular lattice plane.
At low fields, the spins within a single triangular lattice plane
form a 120$^{\rm o}$ structure and as one moves from one plane to
the next the spins are all rotated through an angle $\delta \phi =
q_c c$, so that the component of the IC wave vector along $\hat {\bf c}$
is $q_c$ \cite{MK07}.  To generate the 120$^{\rm o}$ structure, the
in-plane component of the wave vector must be chosen to be at the
corner, ${\bf X}$, of the Brillouin zone of the triangular
lattice, i.e. ${\bf q} = {\bf X} + q_c \hat {\bf c}$.  Then the symmetry
operations ${\cal O}_i$ which leave the wave vector invariant are
$\cal R$ and ${\cal R}^{-1}$.  ($\cal R$ takes ${\bf
X}$ into a vector equivalent to ${\bf X}$.) We thus end up with a 
one-dimensional irrep $\Gamma$ and its complex conjugate $\Gamma^*$.
The spin distribution is given by\cite{MK07,ABH07a}
\begin{eqnarray}
{\bf S} ({\bf r}) &=& [ \sigma_1(q_z)
(\hat i +i \hat j) + \sigma_2(q_z)
(\hat i - i \hat j)]e^{i {\bf q} \cdot {\bf r}} + {\rm c. \ c.} \ ,
\end{eqnarray}
where $\mu= \exp (2 \pi i/3)$.  The order parameters transform as
\begin{eqnarray}
{\cal R} \sigmav_n (q_z) = \mu^n \sigmav_n (q_z) \ , \ \ \ {\cal I}
\sigmav_n (q_z) = \sigmav_{3-n}(q_z)^* \ .
\label{RFMOEQ} \end{eqnarray}
The magnetic free energy up to order $\sigma^4$ is
\begin{eqnarray}
F &=& (T-T_c)\sigmav^2 + u \sigmav^4 + v |\sigma_1(q_z) \sigma_2 (q_z)|^2 \ ,
\label{FMRFMO} \end{eqnarray} where $\sigmav^2 \equiv |\sigma_1(q_z)|^2 +
|\sigma_2(q_z)|^2$, and $u$ and $v$ are constants (with $u$ positive).
It is found\cite{MK07} that only one of the two order parameters is nonzero
in a single domain, from which we deduce that $v$ must be positive. (This
conclusion is confirmed by the appearance of ferroelectricity, as we will
see in a moment.)

\subsection{Magnetoelectric Interaction}

Here we describe the ME interaction which leads to a spontaneous polarization
induced by magnetic ordering which breaks inversion symmetry. For this
purpose we show the dielectric phase diagrams of the three systems under
consideration in figure \ref{PDF}.

We write the free energy as
\begin{eqnarray}
F &=& F_M + F_E + V_{\rm int} \ ,
\end{eqnarray}
where $F_M$ ($F_E$) is the magnetic (dielectric) free energy and $V_{\rm int}$
is the ME interaction which is responsible for the magnetically induced
ferroelectricity.

We first consider NVO\cite{GL05,MK06} and TbMnO$_3$\cite{MK05,ABH07a}. Both have two
magnetic ordered phases, the high-temperature incommensurate
(HTI) phase at higher temperature  ($T_> > T> T_<$), described by a single
order parameter $\sigma_{\rm HTI}$ for which spins are predominantly confined to the easiest direction,
and the low-temperature incommensurate
(LTI) phase (for $T< T_<$) in which a new order parameter $\sigma_{\rm LTI}$ appears, describing ordering transverse
to that of $\sigma_{\rm HTI}$.  The order
parameters are out of phase (to minimize the fourth order terms in the
magnetic free energy)\cite{MK06}, and thus give rise to a magnetic spiral.
These order parameters transform as specified by (\ref{NVOEQ}) and
(\ref{TMOEQ}), respectively.

We have $F_E = (1/2) \chi_E^{-1} {\bf P}^2$, where $\chi_E$ is the dielectric
susceptibility and ${\bf P}$ is the polarization vector. Because there is
no tendency for ferroelectricity to form in the absence of magnetic ordering,
$\chi_E$ never gets large.  In the absence of ME coupling, the equilibrium
value of ${\bf P}$ is zero. The ME interaction has to conserve wave
vector and be invariant under time reversal.  At lowest (quadratic) order in
$\sigma$, it therefore must be of the form $V_{\rm int} \sim \sigma(q) \sigma(-q) P \equiv \sigma \sigma^* P$.
In the present situation, the two $\sigma$'s can
not both be HTI or LTI, because then $V_{\rm int}$ would not be invariant under spatial inversion.  So
\begin{eqnarray}
V_{\rm int} = \sum_\gamma [c_\gamma \sigma_{\rm HTI}(q) \sigma_{\rm LTI}(q)^*
+c_\gamma^* \sigma_{\rm HTI}(q)^* \sigma_{\rm LTI}(q) ] P_\gamma \ ,
\label{TWOOP} \end{eqnarray}
and to be invariant under inversion we must have $c_\gamma=ir_\gamma$,
where $r_\gamma$ is real, so that\cite{GL05}
\begin{eqnarray}
V_{\rm int} &=& i\sum_\gamma r_\gamma [\sigma_{\rm HTI}(q)
\sigma_{\rm LTI}(q)^* - \sigma_{\rm HTI}(q)^* \sigma_{\rm LTI}(q) ]
P_\gamma \label{E12}
\\ &=& 2 \sin(\phi_{\rm HTI} - \phi_{\rm LTI}) |\sigma_{\rm HTI}
\sigma_{\rm LTI}| \sum_\gamma r_\gamma P_\gamma \ , \label{METWO}
\end{eqnarray}
where $\sigmav_{\rm HTI}=|\sigmav_{\rm HTI}|\exp(i \phi_{\rm
HTI})$ and similarly for $\sigmav_{\rm LTI}$.  The transformation
properties given in (\ref{NVOEQ}) and (\ref{TMOEQ}) for the order
parameters under the mirror and glide operations then imply that
$r_\gamma$ in (\ref{E12}) is only nonzero for $\gamma=b$ for
NVO\cite{GL05,MK06,ABH06b,ABH07a} and $\gamma=c$ for
TbMnO$_3$\cite{MK05,ABH06b,ABH07a}. The fact that $P$ is
proportional to $|\sigma_{\rm HTI} \sigma_{\rm LTI}|$ has been
experimentally verified for NVO\cite{FOCUS}.

For RFMO the argument is slightly different. There (\ref{RFMOEQ})
indicates that $\sigma_1(q_z) \sigma_2(q_z)^*$ is invariant under
inversion (which changes the sign of ${\bf P}$).  Thus
(\ref{RFMOEQ}) implies that the ME interaction quadratic in
$\sigma$, which conserves wave vector, is\cite{MK07,ABH07a}
\begin{eqnarray}
V_{\rm int} = \sum_\gamma r_\gamma [ |\sigma_1(q_z)|^2 - |\sigma_2(q_z)|^2]
P_\gamma \ ,
\label{PRMO} \end{eqnarray}
where $r_\gamma$ is real valued.  Since the square bracket is invariant
under the three-fold rotation ${\cal R}$, $P_\gamma$ must also be invariant
under ${\cal R}$. So at this order $r_\gamma$ can only be nonzero for
$\gamma=c$, as is observed\cite{MK07}. At higher order\cite{IS06} a
transverse polarization is in principle possible.  Note that
${\cal R}(P_x-iP_y)=\mu (P_x-iP_y)$ and ${\cal R} \sigma_1 \sigma_2^*
= \mu^2 \sigma_1 \sigma_2^*$. Then one can have an ME interaction of the form
\begin{eqnarray}
V_{\rm int}^{(4)} = c [ |\sigma_1(q_z)|^2 - |\sigma_2(q_z)|^2]
\sigma_1(q_z) \sigma_2(q_z)^* (P_x -i P_y) + {\rm c. c.} \ .
\label{HPRMO}
\end{eqnarray}
However, the fourth order terms written in (\ref{FMRFMO}) select
$|\sigma_1(q_z)|=|\sigma_2(q_z)|$ if $v$ is negative and
$\sigma_1(q_z)\sigma_2(q_z)=0$ if $v$ is positive.
In either case $V^{(4)}_{\rm int}$ does not come into play.
Since the ordered phase is ferroelectric, we deduce that $v$ is
positive and that only $P_c$ is nonzero.  Then, within mean
field theory, $P_c$ is proportional to $\langle |\sigma|^2 \rangle$,
as is the intensity of the magnetic Bragg peaks. This is
experimentally confirmed\cite{MK07}.
\footnote{However, critical fluctuations may imply different exponents for
$P_c$ and $|\sigma|^2$, see Sec. \ref{CP} below.}

\section{125's}


\vspace{-0.3 in}

\begin{figure}[ht]
\hspace{-0.75 in}
\caption{(Color online) \label{125FIG} (a) Two views of the lattice
structure of the 125's. (b)  Symmetry  operations of space group Pbam.
$m_\alpha$ denotes a mirror or glide operation, $2_\alpha$ is
a two-fold rotation or screw operation and $r+\equiv r+1/2$.}

\hspace{-0.75 in}
\includegraphics[width=8.5 cm]{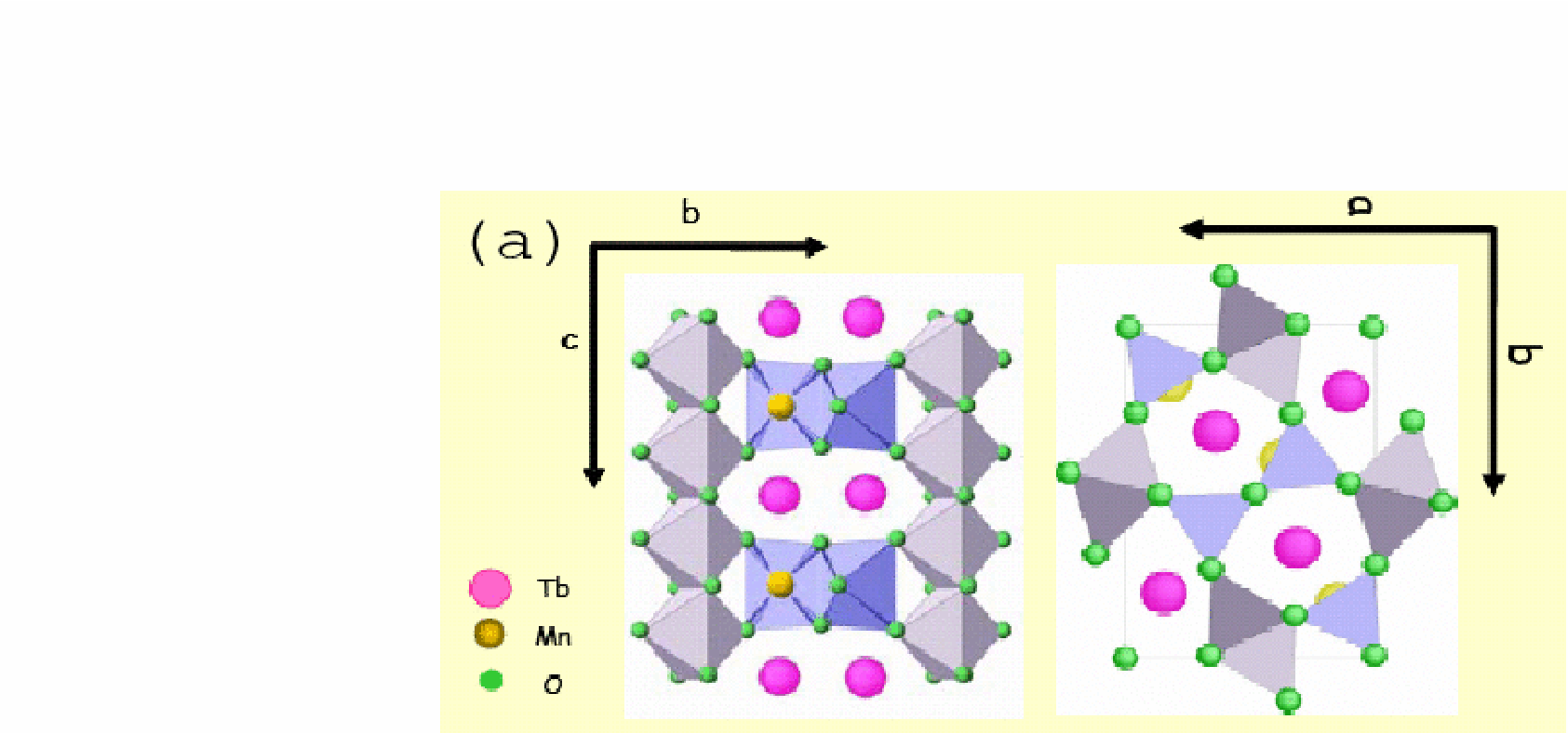}
\hspace{0.2 cm} \includegraphics[width=5.5 cm]{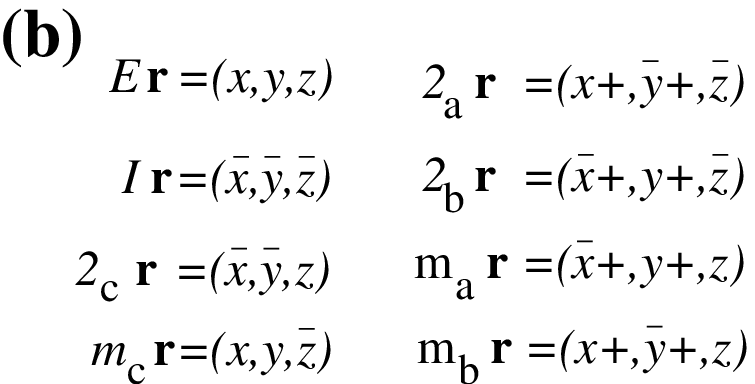}
\end{figure}

\begin{figure}[ht]
\begin{center}
\caption{\label{ETYPD}  ME phase diagrams of
ErMO\cite{SK04a}, TmMO\cite{SK05}, YMO\cite{SK04c,LCC06}, HoMO\cite{DH05,HK06},
TbMO\cite{SK04b}, and DyMO\cite{WR05,RAE08,DH04,NH04a}.  We do not
indicate possible phase changes which have a dielectric signature but
only a weak magnetic signature and hence may represent a minor spin
reorientation.  In Sec. 3.4 we argue that for $40 <
T < 44$K DyMO is in an $(I,I)$ phase.}
\end{center}
\includegraphics[width=5.5cm]{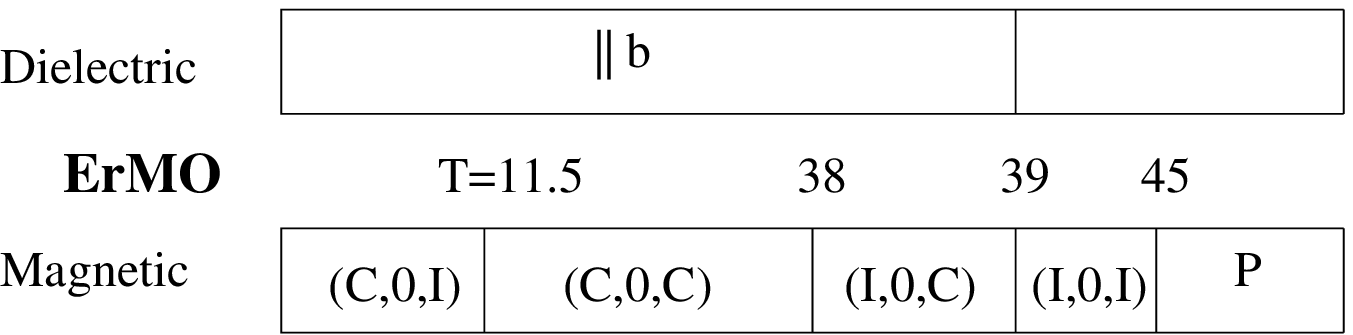}
\hspace{0.75 in} \includegraphics[width=5.5cm]{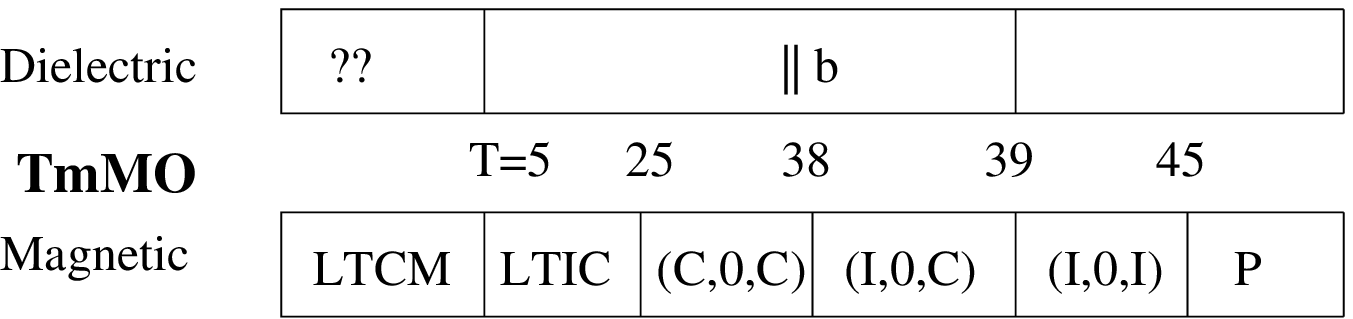}
\vspace{-0.06 in}

\includegraphics[width=5.5cm]{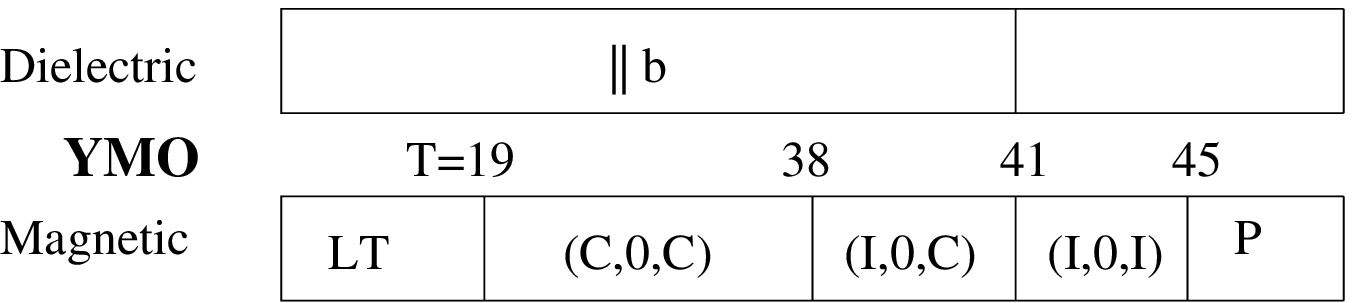} \hspace{0.75 in}
\includegraphics[width=5.5cm]{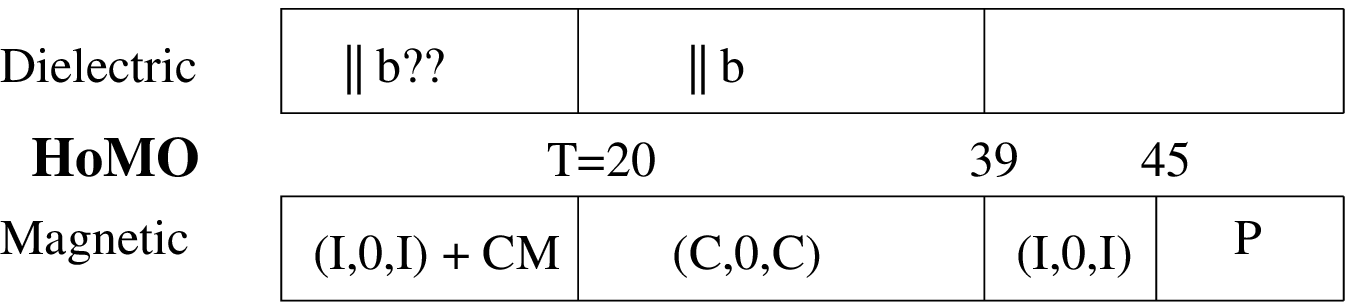} \vspace{-0.06 in}

\includegraphics[width=5.5cm]{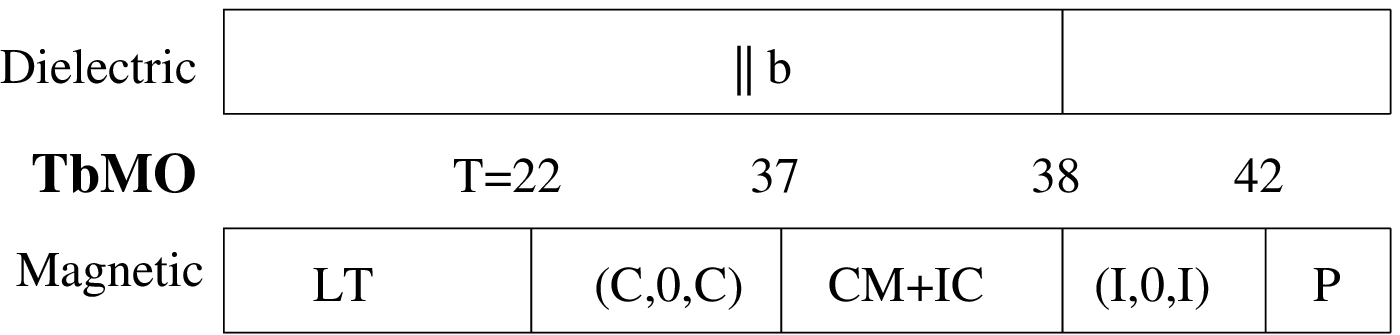} \hspace{0.75 in}
\includegraphics[width=5.5cm]{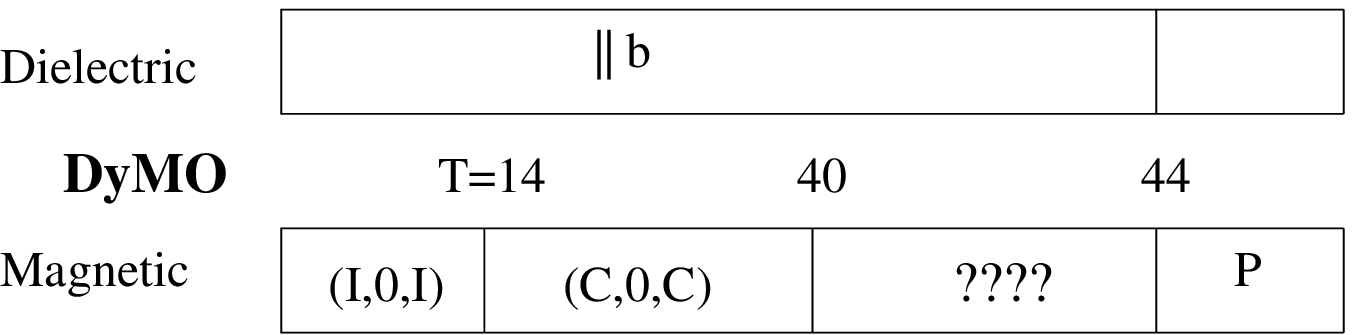}
\end{figure}

\begin{figure}[h]
\begin{center}
\caption{(Color) \label{ICFIG} Left: Schematic 3D phase diagram
for ${\bf q}$ near $(1/2,0,1/4)$.  The top (red) surface
represents the phase boundary between the P and $(I,I)_1$ phases
(where both $q_x$ and $q_z$ are incommensurate). Below the blue
surface, which is a parabola in $J_z$ (depending only weakly on
$J_x$), one has $q_x=1/4$, in phases $(I,C)_1$ and $(I,C)_2$.  The
green surface represents $(I,I)_1 \rightarrow (I,I)_2$ and
$(I,C)_1 \rightarrow (I,C)_2$ (the subscripts 1 and 2 denote the
number of 1D irreps which order). Below the orange surface, which
is a parabola in $J_x$ (depending only weakly on $J_z$), one has
$q_x=1/2$. Right: A cut at constant $q_x$. The $(I,I)_2$ and
$(I,C)_2$ phases disappear below the orange surface (as $q_x
\rightarrow 1/2$), where one has a 2D irrep. The dashed and dotted
lines are possible trajectories followed by specific RMO's as the
temperature is varied.} \vspace{0.2 in}

\includegraphics[width=5. cm]{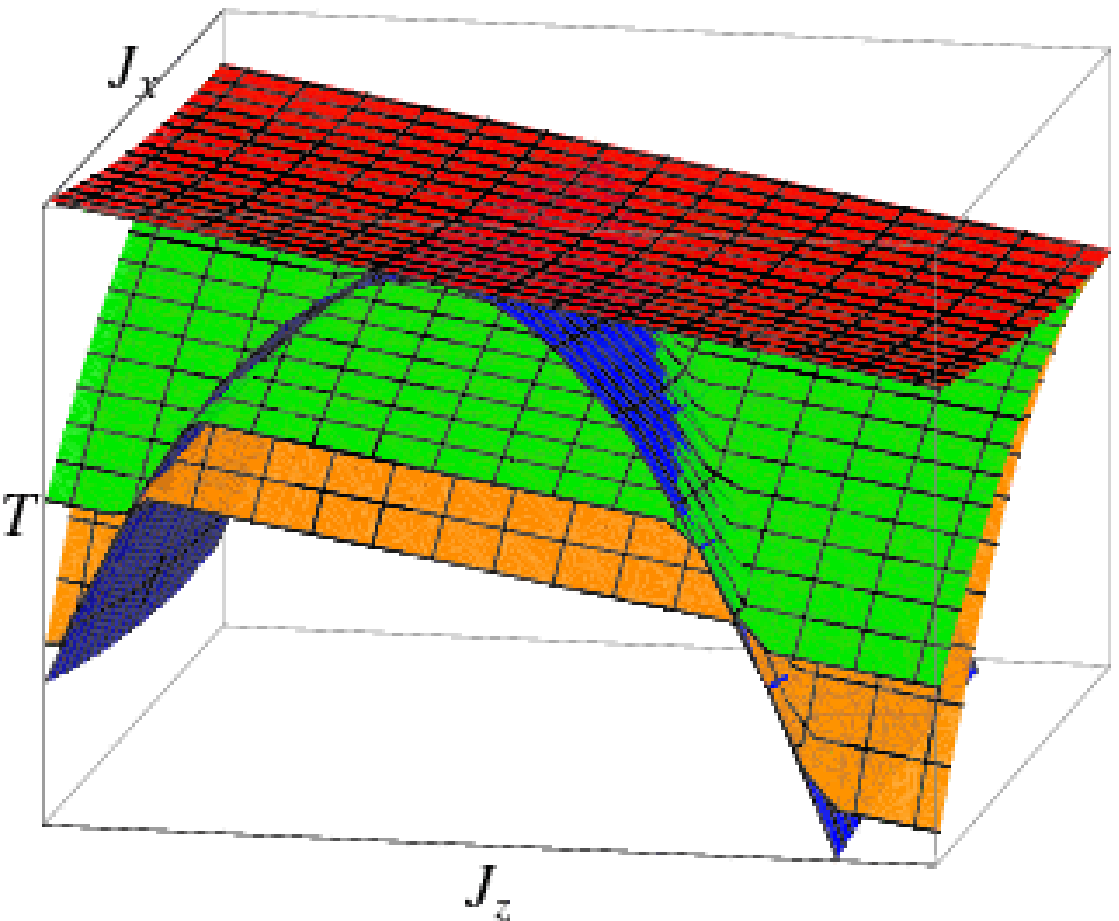}\hspace{1cm}
\includegraphics[width=5.0 cm]{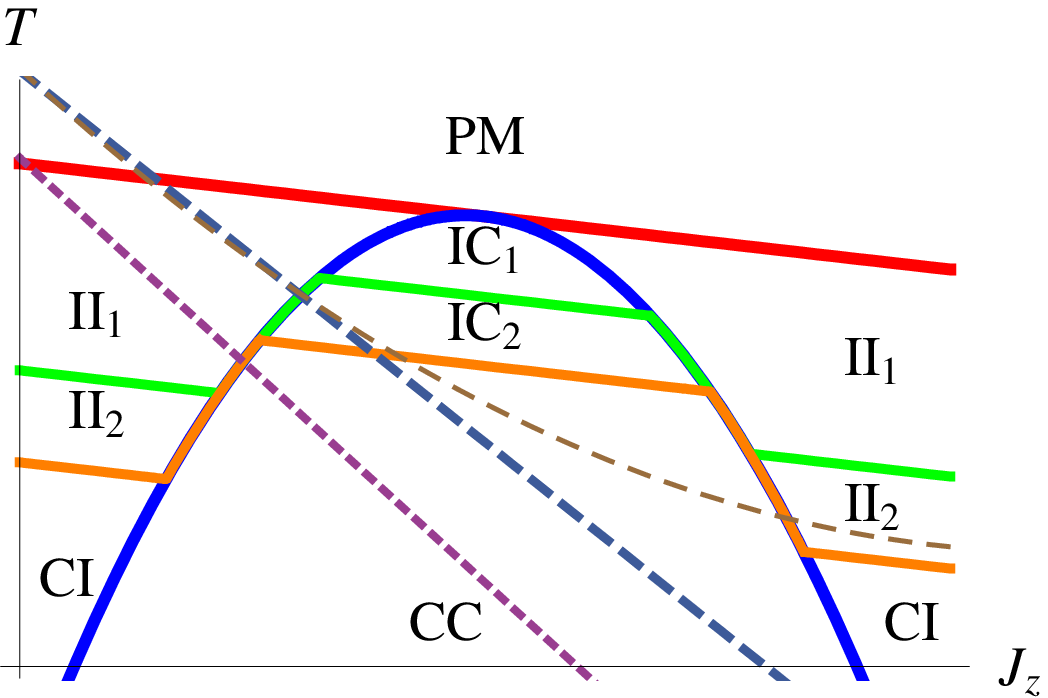}
\end{center}
\vspace{-0.2 in}
\end{figure}

We now consider the ``125" orthorhombic (space group Pbam) family
RMn$_2$O$_5$ (RMO), where R=Y, Ho, Er, Dy, Tb, Tm.  Their lattice
structure and the corresponding space group operations are shown
in figure \ref{125FIG}. The paramagnetic unit cell of the RMO's
contains 12 potentially magnetic
ions: 4 Mn$^{3+}$, 4 Mn$^{4+}$ and 4 R$^{3+}$.
Experiments show that all the RMO's exhibit magnetic spin density
wave ordering, with a wave vector {\bf q} which undergoes a
sequence of phase transitions\cite{LCC04}-\cite{HK06}.
To discuss these phases we introduce the notation ${\bf
q}=(U,0,V)_n$, which we abbreviate as $(U,V)_n$ (in figure
\ref{ICFIG} these are denoted by $UV_n$). If $U=C$ ($U=I$), then
$q_x=1/2$ ($q_x=1/2-\delta$) and if $V=C$ ($V=I$), then $q_z=1/4$
($q_z=1/4+\epsilon$), where the wave vector is in reciprocal
lattice units and $\delta$ and $\epsilon$ are of order 0.01 and
depend on temperature. $V=X$ includes the cases when $\epsilon
\not= 0$ and when $\epsilon=0$. The subscript $n$, if it is given,
indicates indicates the number (1 or 2) of OP's, see below. As the
temperature $T$ decreases, all the RMO's (with the possible
exception of R=Dy) first order below $T_c$ ($\approx 45$ K), into
an incommensurate ($I,I$) phase with no ferroelectric (FE) order. For 
YMO (at $T_F=41$K)\cite{AI96,IK03}, ErMO (at $T_F=39$K)\cite{DH05} 
and TmMO (at $T_F=39$K)\cite{MU98}, this paraelectric incommensurate
state gives way to an $(I,C)$ phase and this phase displays a weak
FE moment ${\bf P}$ along the $b$-axis. Below $T_C\sim 37-39$K,
{\bf q} locks into a commensurate (CM) value $(C,C)$ and $P_b$
increases significantly\cite{IK03}. TbMO\cite{AI96,NH04b},
HoMO\cite{NH04a,DH04}, and probably DyMO\cite{NH04a} go directly
from the $(I,I)$ phase into the ferroelectric $(C,C)$ phase. At
lower temperature (about 10-20K) most of the RMO's return to
having some kind of incommensurate order \footnote{This order may
be commensurate but with a large unit cell.}. We will not be
concerned here with these low temperature phases, since their
existence probably depends sensitively on the details of the
spin-spin interactions.  As we shall see, the behavior of the
higher temperature phases can be described by a generic Landau
free energy.  The magneto-dielectric phase diagrams of various
125's are shown in figure \ref{ETYPD}.

Based on the symmetry of the OP's we construct a Landau theory for
the various RMO's, which yields a generic phase diagram, shown in
figure \ref{ICFIG}, which is {\it independent of the detailed
microscopic interactions}\cite{ABH08a}.
Each RMO has particular coupling constants which
determine the wave vector {\bf q}. Varying these parameters, $J_x$
for $q_x$ and $J_z$ for $q_z$, changes the value of the
optimal ${\bf q}$ at which magnetic ordering occurs. The rest
of this section is devoted to an explanation of this phase diagram
(including the definitions of the various phases) and to a
discussion of its consequences. This analysis is particularly
relevant for the RMO's, because the microscopic theory of their
multiferroicity is somewhat controversial. Our theory provides a
unified explanation for the various sequences of phase transitions
of the magnetic wave vector, and explains why ferroelectricity
does or does not occur in the various magnetic phases. It also
explains the occurrence of two distinct spin structures from
neutron diffraction studies of the CM phase\cite{LCC06,HK07}. This
phenomenological theory suggests several new experiments and makes
a number of predictions, which can be tested experimentally.

\subsection{Magnetic Structure of the $(I,I)_1$ and $(I,C)_1$ Phases of the 125's}

Given the experimental information, we now analyze the various
phases in the order in which they arise upon cooling from the P
phase. The first phase which is encountered is of the $(I,I)$
type. Since $q_z=1/4$ plays no special symmetry role, it is
convenient to discuss the $(I,I)$ and the $(I,C)$ phases together.
 Here,
the star of ${\bf q}$ consists of four wave vectors, namely, ${\bf
q}_\pm = \bigl (\pm(1/2 - \delta), 0, 1/4+\beta \bigr )$ and their
negatives. Each wave vector is invariant under unity and $m_y$.
This symmetry group has two one-dimensional (1D) irreps,
$\Gamma_a$ and $\Gamma_b$, with complex OP's as amplitudes.  By
symmetry, all the wave vectors of the star must have degenerate
eigenvalues of (\ref{F2EQ}).  Therefore, we introduce complex
OP's, $\sigma_a^+\equiv \sigma^{}_a({\bf q}_+^{(a)})$ and
$\sigma_a^-\equiv \sigma^{}_a({\bf q}_-^{(a)})$ associated with
irrep $\Gamma_a$ at its wave vectors ${\bf q}_\pm^{(a)}$, and
similarly for $\Gamma_b$. Here ${\bf q}_+^{(a)}$ and ${\bf
q}_-^{(a)}$ (${\bf q}_+^{(b)}$ and ${\bf q}_-^{(b)}$) are defined
to be the wave vectors at which the $\langle \sigma_a({\bf q})
\sigma_a({\bf q})^*\rangle$ ($\langle \sigma_b({\bf q})
\sigma_b({\bf q})^*\rangle$) susceptibility is maximal as $T
\rightarrow T_{ca}$ ($T \rightarrow T_{cb}$). Specific basis
functions are given elsewhere\cite{ABH08b}, where it is also shown
that they transform as
\begin{eqnarray}
\fl
m_y \sigma^{}_s({\bf q}_\pm^{(s)}) &=& \lambda_s \sigma^{}_s({\bf q}_\pm^{(s)})
, \ \ \ {\cal I} \sigma^{}_s({\bf q}_\pm) =
\kappa_\pm \sigma_s({\bf q}_\pm^{(s)})^* , \ \ \
2_c \sigma_s({\bf q_\pm^{(s)}}) = \eta^2 \sigma_s (-{\bf q}_\mp^{(s)}) ,
\label{eq3}
\end{eqnarray}
where $\lambda_a=-\lambda_b=\exp(i \pi q_x)\equiv \eta^*$
and $\kappa_\pm =\eta^2 \exp(\mp 2 \pi i q_z)$.

As one cools from the P phase, one must enter a phase described by
a single irrep. Arbitrarily choosing this irrep as $\Gamma_a$, the
corresponding free energy is
\begin{eqnarray} F^{(a)}&=&(T-T_{ca})[|\sigma_a^+|^2 +
|\sigma_a^-|^2]+c_1[ |\sigma_a^+|^2 +
|\sigma_a^-|^2]^2\nonumber\\
&&+c_2 |\sigma_a^+ \sigma_a^-|^2+c_3[(\sigma_a^+\sigma_a^-)^2+{\rm
c.c.}]\delta_{4q_z,1}, \label{F4a}
\end{eqnarray}
and analogously for $\Gamma_b$.  The coefficients $c_1,~c_2$ and
$c_3$ may differ for $F^{(b)}$, and we assume that $T_{cb} <
T_{ca}$.  When $q_z\ne 1/4$, this free energy describes the
$(I,I)_1$ phase (the subscript 1 indicates a single irrep). In
this phase, we have $|\sigma_a^+|=|\sigma_a^-|$ if $c_2<0$, while
only one of $\sigma_a^+$ or $\sigma_a^-$ orders if $c_2>0$.
Replacing $(T-T_{ca})$ by $r(q_0)$, where $q_0$ is the wave vector
which minimizes $r(q)$, the corresponding minimal free energies in
the $(I,I)_1$ phase are given by $F_{II}=-r(q_0)^2/w$, with
$w=4c_1$ $(4c_1+c_2)$ if $c_2>0$ $(c_2<0)$.

\begin{figure}[ht]
\begin{center}

\caption{(Color online) \label{CMPDFIG} Phase diagrams (a) for
$q_z \approx 1/4$, based on (\ref{JVSTEQ}) and (b) for $q_x
\approx 1/2$, based on (31) and (32),
when the (I,I)$_1$-(I,I)$_2$ phase boundary (dashed
line) is preempted by locking $q_x$ to $q_x=1/2$. Parabolas shown
as a function of $J_x$ ($J_z$) are weak functions of $J_z$
($J_x$). (c) $r_\pm(q_x)$ for nonzero $\Delta J_x$, based on (30).
The OP associated with each point is given in the box along with 
the parameters which characterize the wave function, as explained 
in \cite{ABH08b}. In (a) and (b) the points M and M' are
multicritical points that can only be reached by adjusting both
the temperature and some additional control parameter.}
\vspace{0.2 in}

\includegraphics[width=3.7 cm]{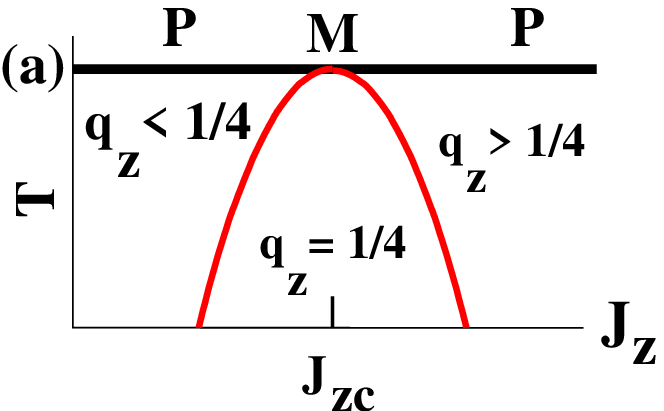}
\hspace{0.05 in}
\includegraphics[width=3.9 cm]{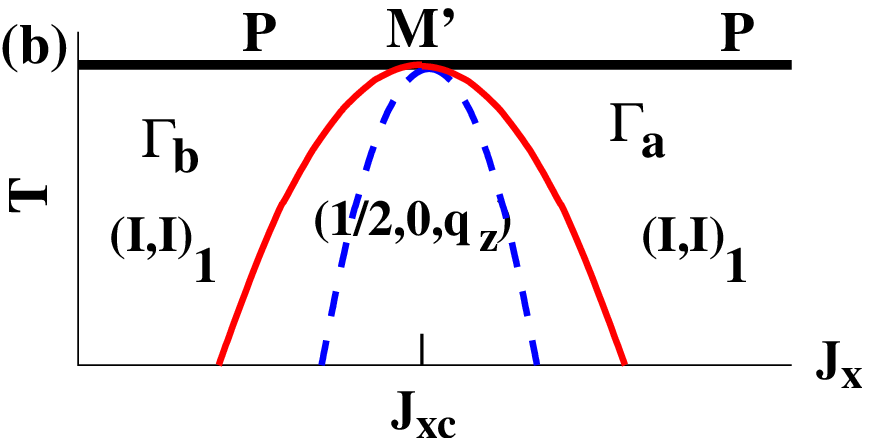}
\hspace{0.05 in}
\includegraphics[width=4.4 cm]{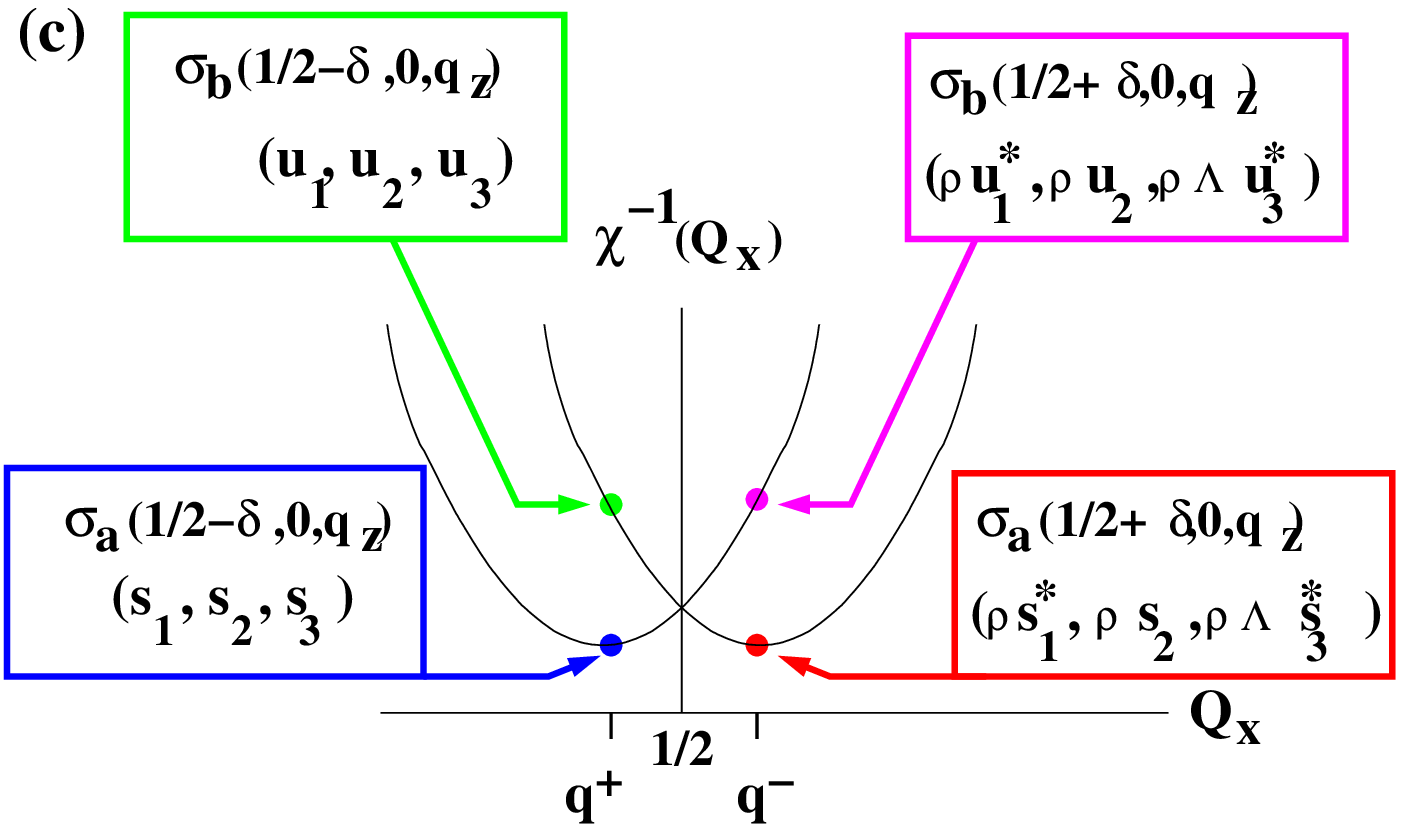}
\end{center}
\label{FIG5}
\vspace{-0.2 in}
\end{figure}

If $q_z$ is close to $1/4$ then the last ({\it Umklapp}) term in (\ref{F4a}) can
lock $q_z$ to $1/4$, via a weakly first order transition. Clearly,
this term arises only when {\it both} $\sigma_a^+$ {\it and}
$\sigma_a^-$ order, which would now happen only if $c_2-2|c_3|<0$.
In this case, one again has $|\sigma_a^+|=|\sigma_a^-|$ and
$F_{IC}=-r(1/4)^2/w'$, with $w'=4c_1+c_2-2|c_3|$. One would then
have a first order transition from $(I,I)_1$ into $(I,C)_1$ when
$F_{II}=F_{IC}$. Since $r(q)$ has a minimum at $q_0$, we have
$r(1/4)\approx r(q_0)+\alpha(1/4-q_0)^2$. Thus, the transition
would occur when $r(q_0)+\alpha(1/4-q_0)^2=r(q_0)(w'/w)^{1/2}$.
Remembering that $r(q_0)=T-T_{ca}$, we have
$q_z-1/4\propto(T_{ca}-T)^{1/2}$. Furthermore, since $q_z=1/4$ is
not a special point, we expect $q_0$ to be a linear function of
$J_z$, hence $1/4-q_0 \propto J_z-J_{zc}$, where $J_{zc}$ is the
special value of $J_z$ associated with the transition from the P
state into the state with $q_z=1/4$. Thus, the transition from
$(I,I)_1$ into $(I,C)_1$ occurs at $T=T_F$, with \begin{eqnarray} T_{ca}-T_F
\propto (J_z-J_{zc})^2, \label{JVSTEQ}\end{eqnarray}
 as shown in figure
\ref{CMPDFIG}a. This parabolic relation is a mean-field result.

Now consider the implications of having $q_z$ locked to the value
1/4 in the $(I,C)$ phases. From (\ref{F4a}) we see that for this
locking to occur, both wave vectors ${\bf q}_+$ and ${\bf q}_-$
must appear (in the same domain). 
Then, since we do not allow a direct transition from the P phase to the
$(I,C)_1$ phase (we ignore the unlikely case of a multicritical
point, at which $J_z=J_{zc}$), the question is whether or not a single
domain of the neighboring $(I,I)_1$ phase has two wave vectors.  From
(\ref{F4a}), the condition to have two wave vectors is that $c_2
<0$. An alternate scenario would be that $c_2>0$ and the two wave
vectors do {\it not} order simultaneously (in the same domain) in
the $(I,I)_1$ phase. In that case, barring the existence of an
as-yet-undetected phase boundary, the two wave vectors would have
to appear in conjunction with the phase transition between the
$(I,I)_1$ and $(I,C)_1$ phases. For the two wave vectors not to be
present in the $(I,I)_1$ phase would imply that $c_2 >0$. Then if
$c_2-2|c_3|<0$, the two wave vectors would appear at the $(I,I)_1
\rightarrow (I,C)_1$ phase transition. It would be interesting to
experimentally determine 
(following the logic of Ref. \cite{SK93})
which scenario actually occurs, i. e.
whether or not the $(I,I)_1$ phase has simultaneous condensation
at both wave vectors. For this purpose, it would be interesting to
perform an experiment analogous to that of Ref. \cite{SK93}.
Here, since the ME interaction is present, one could use an
electric field parallel to one of the wave vectors to
manipulate the domains.

\subsection{Magnetoelectric Structure of the $(I,I)_2$ and $(I,C)_2$ Phases of the 125's}

Similarly to NVO and TbMnO$_3$, the second 1D irrep $\Gamma_b$
may order upon further cooling. In addition to the `decoupled'
free energies $F^{(a)}$ and $F^{(b)}$, the total free energy now
contains many terms which couple the OP's $\sigma_a^\pm$ and
$\sigma_b^\pm$. We start by discussing the $(I,I)_2$ phase,
where all the wave vector components remain incommensurate. 
Then the quartic terms which couple the two sets of OP's are
given by
\begin{eqnarray}
 F_4^{(x)} &=& c_4 [|\sigma_a^+ \sigma_b^+|^2
+ |\sigma_a^- \sigma_b^-|^2]+c_5 [|\sigma_a^+ \sigma_b^-|^2+
|\sigma_a^- \sigma_b^+|^2] + \Delta\ , \label{EQAB}
\end{eqnarray}
where $\Delta$ contains the locking terms,
\begin{eqnarray}
\Delta &=& c_6[ (\sigma_a^+ \sigma_b^{+*})^2 + (\sigma_a^-
\sigma_b^{-*})^2 + (\sigma_b^+ \sigma_a^{+*})^2 + (\sigma_b^-
\sigma_a^{-*})^2] \delta_{{\bf q}_+^{(a)},{\bf q}_+^{(b)}}
\nonumber \\ && + c_7[ \sigma_a^+ \sigma_a^-(\sigma_b^+
\sigma_b^-)^* +  (\sigma_a^+ \sigma_a^-)^* \sigma_b^+ \sigma_b^-]
\delta_{ q_{+,z}^{(a)}, q_{+,z}^{(b)}} \nonumber \\
&& + c_8[ \sigma_a^+ \sigma_b^-(\sigma_b^+ \sigma_a^-)^* +
(\sigma_a^+ \sigma_b^-)^* \sigma_b^+ \sigma_a^-] \delta_{
q_{+,x}^{(a)}, q_{+,x}^{(b)}} \ .\label{Delta}
\end{eqnarray}

Notice that so far we have not assumed that ${\bf q}_\pm^{(a)}$
and ${\bf q}_\pm^{(b)}$ are identical. If the exchange
interactions were isotropic, then the inverse susceptibility would
be invariant under a global rotation of all spin directions. Here,
in the generic case, we have small anisotropic interactions which
break this degeneracy and, in principle, would cause these
critical wave vectors to be slightly different. In this case, the
quartic terms $\Delta$ can lock the wave vectors of the two modes
into equality, as happens for NVO\cite{MK06}. The mechanism for
this locking is as follows. Assume that, say, $\sigma_a$ orders
first at $T=T_{ca}$, and for simplicity we first treat the case with
only a single wave vector, so that, say, $\langle \sigma_a^+
\rangle$, but not $\langle \sigma_a^- \rangle$, is nonzero.
 In analogy with what happens for NVO\cite{MK06},
we assume that ${\bf q}_+^{(a)}$ is almost equal to  ${\bf
q}_+^{(b)}$, at which the inverse susceptibility $\chi^{-1}_b({\bf
q})$ of $\sigma_b$ has its minimum. For $T_{ca}>T>T_{cb}$ this
minimum in $\chi_b^{-1}({\bf q})$ is positive since $\sigma_b$ has
not yet ordered.  Now, the quartic terms $\Delta$ give rise to an
effective quadratic term, $V_{2, {\rm eff}}$. Since only $\langle
\sigma_a^+ \rangle$ is nonzero, we have
\begin{eqnarray}
V_{2,{\rm eff}} &=& c_6[ \langle \sigma_a^+ \rangle^2(
\sigma_b^{+*})^2
+ (\sigma_b^+)^2 \langle \sigma_a^{+*} \rangle^2 ]
\delta_{{\bf q}_+^{(a)},{\bf q}_+^{(b)}} \ , \label{V2EFFEQ}
\end{eqnarray} where $\langle X \rangle$ indicates the thermal
average of $X$. {\it Even before $\sigma_b^+$ orders}, this term
gives an additional contribution [beyond $(T-T_{cb})$] to the
inverse susceptibility of $\sigma_b^+$, but only when ${\bf
q}_+^{(a)}={\bf q}_+^{(b)}$. Since this additional term depends
on the relative phase of the $\sigma_a$'s and the $\sigma_b$'s,
the minimization of this term fixes the phase of $\sigma_b^+$,
reducing its symmetry from that of the XY model (two
components of a complex number) to that of an Ising model.
The minimzation always leads to a negative contribution to
the inverse susceptibility of $\sigma_b^+$.
If $|\langle \sigma_a^+\rangle|^2$ is sufficiently large, this
term can thereby shift the minimum in the $\sigma_b^+$ inverse
susceptibility from the wave vector ${\bf q}_+^{(b)}$ (which it
would have had when $\Delta=0$) into equality with ${\bf q}_+^{(a)}$.
Also, the star of the wave vector associated with $\sigma_b^+$ now 
contains only the two vectors ${\bf q}_+^{(a)}$ and $- {\bf q}_+^{(a)}$.
This scenario applies if the wave vectors for $\sigma_a$ and
$\sigma_b$ are close enough to be locked to ${\bf q}_+^{(a)}$ by
the term $V_{2, {\rm eff}}$ before reaching the temperature
$T_{cb}$ at which $\sigma_b$ condenses. This, in turn, relies on
the smallness of the anisotropic terms which cause ${\bf
q}_+^{(a)}$ to differ from ${\bf q}_+^{(b)}$.

If both $\sigma_a({\bf q}_+^{(a)})$ and $\sigma_a({\bf
q}_-^{(a)})$ condense at $T=T_{ca}$, then we need to consider all
the terms in (\ref{Delta}).  In the $(I,I)_1$ phase, both $\langle
\sigma_a^+\rangle=xe^{i\phi}$ and $\langle
\sigma_a^-\rangle=xe^{i\chi}$ break the symmetry and have well
defined phases $\phi$ and $\chi$ ($x$ is a real number).
Substituting these values into (\ref{Delta}) the yields a
quadratic form in the four real and imaginary parts of
$e^{i\phi}\sigma_b^+$ and $e^{i\chi}\sigma_b^-$, with eigenvalues
$2x^2[c_6\pm (c_7+c_8)]$ and $2x^2[c_6\pm(c_7-c_8)]$. Since only
one of these eigenvalues is lowest, only one combination of the
four OP components of $\sigma_b^\pm$ orders, and thus we still have an
Ising-like ordering into $(I,C)_2$. In any case,  we henceforth
assume that both OP's have the same critical wave vectors.

Experimentally, it seems that the phase $(I,I)_2$ has never been
observed. Instead, the phase with two OP's below $(I,I)_1$ seems
to be of the $(I,C)$ kind. Therefore, we now consider the possible
locking of $q_z$ to $1/4$, which would correspond to the
appearance of the $(I,C)_2$ phase. When $q_z$ is close to $1/4$,
(\ref{EQAB}) must include additional
 {\it Umklapp}
terms, which are also consistent with the symmetry of (\ref{eq3})
and which lock $q_z$ to 1/4. For $q_x \not= 1/2$, these
are
\begin{eqnarray}
 U_{ab} &=& \{c_9\sigma_a^+\sigma_b^+\sigma_a^-\sigma_b^-+c_{10}[(\sigma_a^+\sigma_b^-)^2+\sigma_a^-\sigma_b^+)^2]+{\rm c.c.}\}\delta_{4q_z,1} ,
 \label{UEQ} \end{eqnarray} where
$c_9$ and $c_{10}$ are real. The locking is stronger when two
irreps, rather than a single irrep as in (\ref{F4a}), are present,
because then the additional terms of (\ref{UEQ}) come into play.
However, in either case, note that this locking requires the
presence of {\it both} wave vectors ${\bf q}_+$ and ${\bf q}_-$.

Finally, we discuss the ME interactions in the $(I,I)$ and $(I,C)$
phases.  In analogy with (\ref{E12}), the lowest order ME interaction
which is invariant under the operations of (\ref{eq3}) is
\begin{eqnarray}
V_{\rm int} &=& ir P_y \sum_\pm [ \sigma_a({\bf q}_\pm)
\sigma_b({\bf q}_\pm)^* - \sigma_a({\bf q}_\pm)^* \sigma_b({\bf
q}_\pm)] \ . \label{MEII} \end{eqnarray} Thus, in the $(I,I)$ and
in the $(I,C)$ phases, at this order, ferroelectricity requires
the presence of two order parameters which are not in phase with
one another. At fourth order in the magnetic order parameters,
the ME interaction can lead to small spontaneous polarizations in
the other coordinate directions, but due to space limitations we
refer the reader to \cite{ABH08a}.

\subsection{Magnetoelectric Structure of the $(C,X)$ Phases}

This case includes both $X=I$ ($q_z=1/4+\epsilon$) and $X=C$
($q_z=1/4$). Because ${\bf q}$ is on the Brillouin zone boundary
($q_x=1/2$), the wave vector is invariant under $m_a$ and $m_b$, and 
the star of ${\bf q}$ consists of ${\bf q}$ and $-{\bf q}$. These
operations lead to a two-dimensional irrep\cite{GRB05,ABH07a} and
we choose the basis functions as in table XVI of \cite{ABH07a}.
The actual wave function is a linear combination of the two basis
functions with complex amplitudes $\sigma_1 ({\bf q})$ and
$\sigma_2 ({\bf q})$. These are the OP's which characterize the
magnetic structure and they transform as\cite{ABH07a}
\begin{eqnarray}
\fl \hspace{0.5 in} m_x \sigma_n({\bf q}) &=& \zeta_n
\sigma_n({\bf q}) \ , \ \ \ m_y \sigma_n({\bf q}) = \zeta_n
\sigma_{3-n}({\bf q}) \ , \ \ \ {\cal I} \sigma_n({\bf q}) =
\sigma_{3-n}({\bf q})^* \ , \label{XYIEQ} \end{eqnarray} where
$\zeta_n \equiv (-1)^{n+1}$.  Consistent with these symmetries the
magnetic free energy up to quartic order in $\sigma$ is
\begin{eqnarray}
&& F_M = (T-T_C) [|\sigma^{}_1 ({\bf q})|^2 +|\sigma^{}_2 ({\bf
q})|^2]
+u [|\sigma^{}_1 ({\bf q})|^2 +|\sigma^{}_2 ({\bf q})|^2]^2 \nonumber \\
&& \ \ + w |\sigma^{}_1({\bf q}) \sigma^{}_2 ({\bf q})|^2 +  v
[\sigma^{}_1({\bf q}) \sigma^{}_2({\bf q})^* +\sigma^{}_2({\bf q})
\sigma^{}_1({\bf q})^* ]^2 \nonumber \\
&& \ \ + [ x (\sigma_1({\bf q})^4 + \sigma_2({\bf q})^4) + y
\sigma_1({\bf q})^2 \sigma_2({\bf q})^2 + {\rm c. \ c.} ]
\delta_{4q_z,1} \ , \label{F4}
\end{eqnarray}
where $x$ and $y$ are real. Under the terms quadratic in $\sigma$ and
those scaled by $u$, all directions in the four dimensional
space of $\sigma_1\equiv \sigma^{}_1({\bf q})$ and $\sigma_2
\equiv \sigma^{}_2({\bf q})$ are equally unstable relative to
ordering.  However, for $q_z \not=1/4$, the fourth order terms
select $|\sigma_1|=|\sigma_2|$ for $w+4v<0$ if $v$ is negative,
$\sigma_1=\pm i \sigma_2$ for $w<0$ if $v$ is positive, and
$\sigma_1\sigma_2=0$ otherwise.  For $q_z=1/4$ the terms in $x$
and $y$ are difficult to analyze analytically, but in many cases
we find that the phases of $\sigma_1$ and of $\sigma_2$ can be
chosen so that $F_M$ still has minima when either
$|\sigma_1|=|\sigma_2|$ or $\sigma_1 \sigma_2=0$.

Now we consider the dielectric properties.  At quadratic order in
$\sigma$, since ${\cal I} \sigma_1 \sigma_2^*=\sigma_1\sigma_2^*$,
(\ref{PRMO}) also applies to the 125's when ${\bf q}=(1/2,0,q_z)$
and then (\ref{XYIEQ}) indicates that $r_\gamma$ is only nonzero
for $\gamma=b$. Including terms of higher order in
$\sigma$\cite{IS06} the ME interaction for the 125's is of the
form
\begin{eqnarray}
\fl \hspace{0.25 in} V_{\rm int} = r_c [ |\sigma_1|^2 -
|\sigma_2|^2] P_b
+ i\sum_\gamma r'_\gamma [ |\sigma_1|^2 - |\sigma_2|^2] [ \sigma_1
\sigma_2^* - \sigma_1^* \sigma_2]P_\gamma \ , \label{125ME}
\end{eqnarray} where, according to (\ref{XYIEQ}), the real
coefficient $r'_\gamma$ is only nonzero for $\gamma=a$. However,
as mentioned above, (\ref{F4}) probably allows only either
$|\sigma_1|=|\sigma_2|$ or $\sigma_1 \sigma_2=0$, in which case
the last term in (\ref{125ME}) is
inoperative. On the other hand, if $\sigma_1 \sigma_2=0$ (so that,
say, $\sigma_2=0$) and if one applies an electric field, $E_a$, in
the ${\bf a}$ direction, which induces a nonzero value of $P_a$,
the second term in (\ref{125ME}) will induce a nonzero
out-of-phase value in the order parameter, $\sigma_2$, that was
zero for $E_a=0$. Then with $\langle \sigma_1 \rangle \not= 0$,
this effective linear coupling between $P_a$ and $\sigma_2$ gives
rise to electromagnons\cite{AP06,RVA1,RVA2}.

The ME coupling can induce lattice displacements at wave vectors
which are even integer multiples of the magnetic wave
vector\cite{TK03,Rad08}. Since the results are particularly simple for
the $(C,C)$ phase, where ${\bf q}=(1/2,0,1/4)$, we now
discuss the lowest order interaction in that case.  So far we
considered a trilinear spin-phonon coupling involving $\sigma({\bf
q}) \sigma ({\bf q})^*$, which conserves wave vector and therefore
couples to a uniform polarization. We now generalize this
analysis, and consider terms of the form $\sigma({\bf q})^2$ or
$\sigma({\bf q})^{*2}$, which couple to phonon modes with wave
vector $\pm 2{\bf q}$. Within a reciprocal lattice vector, this
phonon wave vector is equal to the antiferroelectric wave vector
$(0,0,1/2)$. To construct this interaction we need the site
symmetry analysis for this wave vector, which is the same as for
the wave vector $(0,0,0)$ as given in Table I of  \cite{BM05}.
There it is indicated that there are 15 B$_{3u}$ ($x$-like) modes,
15 B$_{2u}$ ($y$-like) modes, and 9 B$_{1u}$ ($z$-like) phonon
modes. An $x$-like mode, for instance, need not involve
displacements along the $x$-axis; rather such a mode need only
transform like $x$ under the space group operations.  \footnote{As
explained in  \cite{TY06}, the largest polarization will come from
$r_\alpha$-like modes, which have displacements in the
$r_\alpha$-direction.} Accordingly, let $u_{\rm AF}(\gamma,
\tau)$ denote such a phonon, where $\gamma$ labels the symmetry
($x$, $y$ or $z$, since we are only interested in vector-like
modes which carry a polarization) and the index $\tau$ labels the
occurrence. We use the transformation properties of (\ref{XYIEQ})
with $m_z={\cal I}m_xm_y$, so that $m_z \sigma_n = \sigma_n^*$.
Thus the combination $(\sigma_1^2 + \sigma_2^2)$ is even under
$m_x$ and $m_y$, so that the spin-phonon interaction contains the
term
\begin{eqnarray}
V_{{\rm sp-ph},z} &=&  \sum_\tau [ir_\tau(\sigma_1^2 + \sigma_2^2)
+ c. c. ] u_{\rm AF}(z,\tau) \ ,
\end{eqnarray}
where $r_\tau$ is real, so that the square bracket is odd under
$m_z$. Similarly $\sigma_1 \sigma_2$ is odd under $m_x$ and $m_y$,
so it cannot couple to a vector. Finally $(\sigma_1^2-\sigma_2^2)$
is even under $m_x$ and odd under $m_y$ and it gives rise to an ME
interaction of the form
\begin{eqnarray}
V_{{\rm sp-ph},y} = \sum_\tau [r_\tau'(\sigma_1^2 - \sigma_2^2) +
c. c. ] u_{\rm AF}(y,\tau) \ ,
\end{eqnarray}
where  $r_\tau'$ is real and we noted that the square bracket is
even under $m_z$.  In summary, at this order one can have
antiferroelectricity with polarization along either $y$ or $z$.

We next analyze the tongue associated with $q_x=1/2$. Note that for a
critical value, $J_{xc}$, of the control parameter $J_x$, the two
branches [denoted $r_\pm (q_x,J_x)$] of the quadratic coefficients
$r(q_x)$ of the inverse susceptibility
are degenerate and are minimal at $q_x=1/2$, so that
$r_\pm (q_x,J_{xc}) = r(0) + a(q_x-1/2)^2 + {\cal O}(q_x-1/2)^4$,
where $a$ is a positive constant.
As $J_x$ is varied away from $J_{xc}$, a term in $r_\pm(q_x)$ which
is linear in $k_x \equiv (1/2-q_x)$ is allowed and generically
is of order $\Delta J_x \equiv J_x-J_{xc}$\cite{ABH08b}.
\footnote{To see the existence
of such a term consider the approximation in which, for a system
with isotropic exchange interactions $J({\bf r},{\bf r}')$ between
spins at ${\bf r}$ and ${\bf r}'$, one has, for $\tau \not= \tau'$
and ${\bf q}=(1/2-\delta,0,q_z)$ that
$\chi^{-1}_{\tau,\alpha; \tau' \alpha}({\bf q}) = J(\tauv , {\bf a}
+ \tauv')\exp [\pi i -2 \pi i \delta )] + J(\tauv , -{\bf a}
+ \tauv')\exp [-\pi i + 2 \pi i \delta )]$. As long as the
sites do not sit at a center of inversion symmetry, these two terms
will have different amplitudes and will give an imaginary contribution
which is linear in both $\delta$ and $J$.}
The symmetry operation $m_x$ dictates that the spectrum of the
two branches $r_\pm (q_x)$ should be independent of the
sign of $k_x$, as shown in figure \ref{CMPDFIG}c, so that
\begin{eqnarray}
r_\pm(k_x,J_x)&=& r(0)+a k_x^2 \pm b k_x (J_x-J_{xc})\ .
\label{req} \end{eqnarray} and for concreteness we assume that the
constant $b$ is negative and that $J_x > J_{xc}$.  Symmetry thus
implies the existence of equivalent minima at $k_x = \mp
b(J_x-J_{xc})/(2a)\equiv k^\pm$. Thus at its minimum $r_\pm(k_x)$
assumes the value $r(k_x=0) - \alpha' (J_x - J_{xc})^2$, where
$\alpha'$ is a constant. Accordingly, we can adopt the argument
leading to (\ref{JVSTEQ}), to the present case and obtain
\begin{eqnarray}
T_{ca} - T_{C} \propto (J_x - J_{xc})^2 \ ,
\end{eqnarray}
where $T_C$ is the phase boundary between the $(I,X)$ and $(C,X)$
phases. (This phase boundary is the solid line in Fig.
\ref{CMPDFIG}b.)

The structure of (\ref{req}) also allows us to discuss the phase
boundary $T_{1 \rightarrow 2}$ between the $(I,I)_1$ and $(I,I)_2$
phases. For that purpose we compare (\ref{req}) with (\ref{F4a})
(and with its analog for $F^{(b)}$) and identify $r_+$ with
$T-T_{ca}$ and $r_-$ with $T-T_{cb}$. We thereby find that
\begin{eqnarray}
T_{c,a}-T_{c,b}= 2bk^+(J_x-J_{xc}) \sim c (J_x-J_{xc})^2 \ ,
\end{eqnarray}
where $c$ is a constant.  Thus  $T_{1\rightarrow 2}$ is
proportional to $(J_x-J_{xc})^2$. Depending on the parameters,
this parabolic tongue can be either narrower or wider than that
considered above for locking $q_x$ to $q_x=1/2$. In the figure we
show the former case, since the $(I,I)_2$ phase has not been
observed for any of the 125's.

\subsection{Generic Phase Diagram for RMn$_2$O$_5$}

We now explain how the generic phase diagram of figure \ref{ICFIG}
describes the various RMO's.  Since $q_z=1/4$ is not a high
symmetry point, we can not condense from the P phase into
$q_z=1/4$ unless we adjust the $J$'s appropriately to reach this
higher order multicritical point. Since we reject this accidental
possibility, the first ordered phase we encounter has $q_z
\not=1/4$.  Although $q_x=1/2$ {\it is} a special value
(characteristic of antiferromagnetically doubling the size of the
unit cell), the result shown in figure \ref{CMPDFIG}b indicates
that a continuous transition from the P phase into a $(C,I)$ phase
is not allowed because it would also involve a multicritical
point. For the RMO's (except R=Dy which we discuss separately),
experiment shows that the first ordered phase is $(I,I)$ and this
case is shown in figure \ref{ICFIG}. From now on we arbitrarily
set $T_{c,a} > T_{c,b}$ (since we reject the possibility of
accidental equality). Consequently we identify that the transition
from the P phase is into an ordered phase $(I,I)_1$  with a single
OP $\sigma^{}_a$ (except for the star of ${\bf q}$). For a single OP, (\ref{MEII}) provides a
phenomenological explanation for why this phase is not
ferroelectric.  As discussed above, we assume that in the
$(I,I)_1$ the phases ${\bf q}_\pm^{(a)}$ and ${\bf q}_\pm^{(b)}$ become
locked into equality {\it without} crossing a phase boundary. For
the phases with $q_x \not= 1/4$, experiments have not yet
indicated whether the two wave vectors ${\bf q}_{\pm}$ occur in
separate domains, or whether the true state is the superposition,
within a single domain, of the two wave vectors. As $T$ is further
reduced through the $(I,I)_1$ phase, a second continuous
transition could occur, producing a phase $(I,I)_2$ in which both
OP's $\sigma{}_a$ and $\sigma{}_b$ are nonzero (as in
NVO\cite{GL05,ABH07a} or TbMnO$_3$\cite{MK05}).

The above description applies for $J_z$ relatively far away from
$J_{zc}$, i.e. $q_z$ relatively far away from $1/4$. If $q_z=1/4$,
one goes directly from the P phase into the $(I,C)_1$ phase, which
is similar to the $(I,I)_1$ phase. Upon cooling, the OP related to
the other 1D irrep tends to order, and one has a transition into
the $(I,C)_2$ phase. This transition happens at a higher
temperature than that for $(I,I)_1\rightarrow(I,I)_2$, due to {\it
Umklapp} terms like (\ref{UEQ}), which enhance the tendency of
$\sigma_b(q_x,0,\pm 1/4)$ to order (compared to
$\sigma(q_x,0,q_z)$ with an IC $q_z$). If $q_z$ is close to $1/4$,
one first goes from the P phase into the $(I,I)_1$ phase, but then
the {\it Umklapp} terms cause a transition into the $(I,C)_1$
phase, and one ends up with the phase diagram shown on the RHS of
figure \ref{ICFIG}.

As the temperature is lowered, each individual RMO follows some
trajectory in the parameter space. The RHS plot in figure
\ref{ICFIG} shows possible projections of such trajectories. The
trajectories, as well as the optimal wave vectors,
are assumed to have some temperature dependence, which
can originate from the elimination of secondary degrees of
freedom, which generate effective temperature-dependent exchange
coefficients. Note that the whole diagram corresponds to the close
vicinity of ${\bf q}=(1/2,0,1/4)$, so that this temperature
dependence is relatively weak. As shown in figure \ref{ETYPD}, the
real RMO's go directly from the $(I,I)_1$ phase into  either an
$(I,C)$ (for R=Er, Tm, Y) or into the $(C,C)$ phase (for R=Ho, Dy,
Tb).  In the former case, we now argue that this phase must be the
$(I,C)_2$ phase: Since the experimentally observed  phase is
ferroelectric, it follows that there must exist {\it two} OP's,
$\sigma_a$ and $\sigma_b$. Once both order parameters exist, this
phase could be either $(I,I)_2$ or $(I,C)_2$. Since the
experiments find that $q_z=1/4$, this must be $(I,C)_2$. Indeed,
we conclude that the trajectories for R=Er, Tm, Y are represented
by the dashed lines with long dashes in the phase diagram.  As the
same lines indicate, one would then go into the $(C,C)$ phase, as
indeed observed. At lower temperatures, the trajectories could
leave the $(C,C)$ phase to the other side of the parabolic
`tongue', and enter a less commensurate phase, which could be
paraelectric [$(I,I)_1$] or ferroelectric [$(I,I)_2$ or $(C,I)$].

As indicated by the dashed line with short dashes in the same figure, one can also go directly from $(I,I)_1$ into $(C,C)$.
 This trajectory thus describes the RMO's with R=Ho, Dy, Tb. In the $(C,C)$ phase,
which is ferroelectric, (\ref{125ME}) indicates that $|\sigma_1| \not=
|\sigma_2|$.  The quartic term of (\ref{F4}) implies that either
$|\sigma_1| = |\sigma_2|$, or one of them is zero, so that
$\sigma_1 \sigma_2=0$.  Thus only the first term in (\ref{125ME})
survives and it explains the observation\cite{NH04b,NH04a} that the
spontaneous polarization lies along the {\bf b} axis.    Finally, we should
mention that the fact that different R's follow slightly different
trajectories is reasonable from the following qualitative point of view.
For Tm, Er, and Y the value of $q_x$ (listed in table \ref{QZTAB}) is much
closer to $1/4$ and therefore is more likely to be locked to $q_x=1/4$
than is that of Ho and Tb.

\begin{center}
\begin{table}
\caption{\label{QZTAB} Values of $q_x$, $q_z$, and
$\Pi \equiv |1/2-q_x| -|1/4-q_z|$ for $T$ near $T_c$ for various RMO's.
Positive $\Pi$ favors locking $q_x$ to the value 1/4 in preference to
locking $q_x$ to the value 1/2.}

\begin{center}
\vspace{0.2 in}
\begin{tabular} {||c | c c c c c||}
\hline R$=$ & Tm\cite{SK05} & Er\cite{SK04a}& Y\cite{IK01,SK04c} &
Ho\cite{HK06} & Tb\cite{SK04b,LCC04} \\ \hline $q_z=$ & 0.252 & 0.244
&  0.255 & 0.237 & 0.277 \\
$q_x=$ & 0.472 & 0.479 & 0.482 & 0.488 & 0.487 \\
$\Pi =$ & 0.026 & 0.015 & 0.013 & -0.001 & -0.010 \\  \hline
\end{tabular}
\end{center}
\end{table}
\end{center}

For DyMO, experiments have not definitively determined the sequence of
phase transitions in the wave vector, because the large incoherent neutron
cross section of the Dy nucleus causes experimental problems.  A recent
X-ray experiment\cite{RAE08} has confirmed the existence\cite{WR05} of
the $(C,C)$ state. The specific heat\cite{NH04a} provides evidence that
there is a single intermediate phase between this state and the paramagnetic
state. As argued in connection with figure  5, this intermediate phase has
to be an $(I,I)_1$ phase, because we do not allow the possibility of accidentally
hitting the multicritical point where the P phase meets the $(I,C)$
(in figure 5a)  or $(C,I)$ phase (in figure 5b). This proposed phase
exhibits a single OP, which is also consistent with the fact that
DyMO is paraelectric for $T>40$K (see figure 4).

We now return to the phase diagram of figure \ref{ICFIG}. All the
RMO's have ${\bf q}$ close to $(1/2,0,1/4)$ (see table
\ref{QZTAB}), so they leave the P phase near the apex of the
tongue of figure \ref{CMPDFIG}a or \ref{CMPDFIG}b. The effects of
a magnetic field are explained as follows: it generates magnetic
moments on the R ions (even above their ordering temperature).
Since these ions couple to the Mn ions, their moment changes the
effective Mn-Mn interactions, thus changing the 'control
parameters' and the optimal {\bf q}. This often moves the material
towards the $(C,C)$ tongue, resulting in a transition from $(I,C)$
($(I,I)$ when paraelectric) back into the CM
phase\cite{DH05,HK06}. Pressure\cite{CRC07} has similar effects.

\subsection{Spin structures in the $(C,C)$ phase}

\begin{figure}[ht]
\begin{center}
\caption{(Color online) \label{YMOFIG} Schematic diagram of the ${\bf a}$
and ${\bf b}$ components of the Mn$^{3+}$ spins in a single a-b plane
of YMO for the CM phase. 
The glide $m_x$ consists of a mirror plane M at $x=a/4$ followed
by a translation b/2 along $y$.  Left: the structure given in table III
of \protect{\cite{HK07}} (with the ${\bf c}$-components not shown).
Right:  the structure given in figure 2 of \protect{\cite{LCC06}}
(who reported zero \protect{${\bf c}$}-components of spin.)}
\vspace{0.2 in}

\includegraphics[width=8.5 cm]{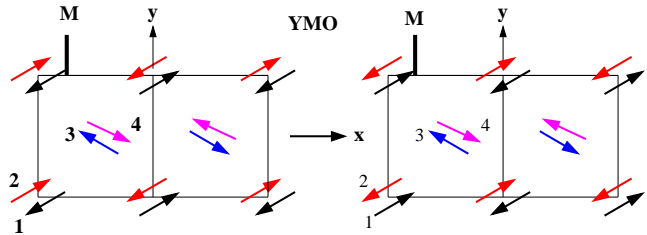}
\end{center}

\end{figure}

The introduction of OP's leads to a natural interpretation of neutron
scattering results for the $(C,C)$  phase in YMO. Figure \ref{YMOFIG} shows the
Mn$^{3+}$ {\bf a}-{\bf b} plane spin components in the CM phase of YMO,
from the neutron diffraction results of \cite{LCC06}
\footnote{The top (bottom) panel of figure 2 in this paper should be
labeled 24.7K (1.9K).} and \cite{HK07}.
These two structures are obviously similar, and one might ask what
symmetry (if any) relates them.  (This degeneracy was also found
in the first-principles calculation of \cite{CW07}.) We now show
that these two structures are indeed equivalent.\cite{ABH08a}
To identify the
symmetry element that relates them note that the structure on the
left is even under the glide operation $m_x$, while that on the right is
odd under $m_x$. (Here one should note that spin, being a pseudovector,
transforms with an additional minus sign under a mirror operation.)
Then (\ref{XYIEQ}) indicates that
the structure on the left has $\sigma^{}_2=0$, whereas that on the
right has $\sigma^{}_1=0$. Going between these two structures
corresponds to a rotation in OP space. This equivalence is easily
understood when OP's are introduced, as done here. Since
either $\sigma_1=0$ or $\sigma_2=0$, we conclude from the discussion below
(\ref{F4}), that $w + 2 v - 2|v|$ is positive and both OP's can not
order simultaneously\cite{ABH07a,ABH08a}. This conclusion supports
that reached above, namely that since the CM phase is ferroelectric,
the fourth order terms in (\ref{F4}) must select $\sigma_1 \sigma_2=0$.

\begin{center}
\begin{table}
\caption{Magnetic structures of YMn$_2$O$_5$ at $T=25$K
showing the spins vectors of the eight Mn$^{3+}$ and the eight
Mn$^{4+}$ sites within the cell $a_0 \times b_0 \times 2c_0$. The
complete magnetic unit cell is found by antiferromagnetically
doubling the cell in both the {\bf a} and {\bf c} directions.}
\label{HKT}
\vspace{0.2 in}

\begin{scriptsize}
\begin{tabular} {|| c c c c c c|| c || c c c c c c||}
\hline \hline
\multicolumn{6} {||c||} {Mn$^{3+}$} & & \multicolumn{6} {|c||} {Mn$^{4+}$} \\
\hline
$M_x$ & $M_y$ & $M_z$ & $M_x$ & $M_y$ & $M_z$ & $n$ &
$M_x$ & $M_y$ & $M_z$ & $M_x$ & $M_y$ & $M_z$ \\
\hline
\multicolumn{3} {||c|} {H. Kimura {\it et al.}}
& \multicolumn{3} {|c||} {This work}  & &
\multicolumn{3}{c|} {H. Kimura {\it et al.}}
& \multicolumn{3} {|c||} {This work} \\
\hline
 -2.02& -0.41& -0.71& -2.04& -0.38& -0.67&
1&  1.74&  0.51&  0.28&  1.72&  0.55&  0.30\\
  2.20& -0.40& -0.24&  2.18& -0.41& -0.26&
2&  1.69& -0.59& -0.32&  1.72& -0.55& -0.30\\
 -2.06&  0.35&  0.63& -2.04&  0.38&  0.67&
3&  0.98& -0.33& -0.69&  0.99& -0.30& -0.63\\
 -2.15& -0.42& -0.29& -2.18& -0.41& -0.26&
4&  1.00&  0.27&  0.57&  0.99&  0.30&  0.63\\
  2.82&  0.58& -0.51&  2.85&  0.53& -0.48&
5& -1.65& -0.46&  0.51& -1.63& -0.50&  0.56\\
 -3.07&  0.55& -0.18& -3.04&  0.57& -0.19&
6& -1.61&  0.55& -0.62& -1.63&  0.50& -0.56\\
  2.87& -0.49&  0.45&  2.85& -0.53&  0.48&
7& -2.12&  0.74& -0.09& -2.15&  0.68& -0.10\\
  2.99&  0.59& -0.21&  3.04&  0.57& -0.19&
8& -2.18& -0.63&  0.09& -2.15& -0.68&  0.10\\
\hline \hline \end{tabular} \end{scriptsize}

\end{table}
\end{center}

To make this identification more quantitative, we consider the
magnetic structure which  H. Kimura {\it et al.}\cite{HK07}
deduced from their neutron diffraction study, which we summarize in
table \ref{HKT}.  Their structure determination was based on an
unrestricted fit, in which no particular symmetry was assumed.
In contrast, our analysis based on representation theory assumes
that the magnetic structure is characterized by the two complex-valued
order parameters $\sigma_1$ and $\sigma_2$, with corresponding spin
wave functions which are given in \cite{ABH07a}, but more
conveniently in table IX of \cite{ABH08b}. 
Since we expect that $\sigma_1\sigma_2=0$, our theory would imply that the spin structure should be fitted with only one OP component.
Indeed, we find that Kimura {\it et al.}'s data can be fitted with $\sigma_2=0$.
Optimizing the parameters of table IX of \cite{ABH08b}
so as to reproduce the spin structure of Kimura {\it et al.}, we
found the optimal structure constants to be
\begin{eqnarray}
{\bf r}_1 &=& (-0.387, -0.072,  0.091i)\ , \ \ \
{\bf r}_2 = ( 0.413,  0.078,  0.036i) \nonumber \\
{\bf z} &=& (0.257+0.049i, -0.081-0.017i, 0.031-0.063i) \ .
\end{eqnarray}
With the normalization $2|{\bf r}_1|^2 + 2|{\bf r}_2|^2
+ 4| {\bf z}|^2=1$, the complex-order parameter was found to be
\begin{eqnarray}
\sigma_1 =  5.2698 +i 7.3691  \ .
\end{eqnarray}
(This complex phase can not be explained by a low order anisotropy
in the complex $\sigma_1$ plane.) From table \ref{HKT} one sees
that the structure assuming the validity of representation theory
is quite close to that of the unrestricted fit of Kimura {\it et
al.}.  The difference between these two structures is that our
version respects the symmetry one would attribute to a structure
having only $\sigma_1$ nonzero.  Thus, in our structure the
magnetic sublattices are related in pairs, whereas in the
structure of \cite{HK07} these sublattices are almost, but not
exactly, related. To characterize the difference between these two
structures, note that $|\sigma_1|\approx 9.1$ gives the square
root of the sum of the squares of the spin amplitudes within the
cell of table \ref{HKT}.  The analogous quantity for the
difference vector between the two structures is 0.23, indicating
that the difference, if real, corresponds to an additional order
parameter whose magnitude is about 2.5\% of $\sigma_1$.  As we
explained, near the high-temperature limit of this phase one can
only have either $|\sigma_1\sigma_2|=0$ or $|\sigma_1|=|\sigma_2|$.
Thus, if $|\sigma_2| \ne 0$ then we would expect it to be of the
same order as $|\sigma_1|$. Thus, it seems unlikely that if such
an additional order parameter would emerge, it would be so small
 deep in the CM phase, where the data were taken.
 Accordingly, we propose that the actual
magnetic structure in the $(C,C)$ phase of YMO corresponds to a
single order parameter $\sigma_1$.  We have also identified that
the data from \cite{HK07} on HoMn$_2$O$_5$ exhibit the same
symmetry: namely the $(C,C)$ phase is characterized by the single
order parameter $\sigma_1$. Similarly, we identify that the
magnetic structure of the Mn spins in ErMn$_2$O$_5$, as reported
in \cite{HK07}, is also consistent with the symmetry associated
with the single order parameter $\sigma_1$. However, the phases
$\phi_x$ of the $x$-components of the Er magnetic moments ($0.8
\pi$ and $-0.3 \pi$) do not agree with the values ($\pi$ or 0)
corresponding to $\sigma_1$. It would be interesting to check the
sensitivity of the data to variation of these phases. It is
interesting that the structures of all the 125's determined in
\cite{HK07} have $\sigma_2=0$, even though the structure with
$\sigma_1=0$ represents an equivalent way that magnetic ordering
can break symmetry. Apparently, the sample preparation (which
might create some uniaxial strain) or some other experimental
detail (stray electric fields?) chooses the structure with
$\sigma_2=0$ in these experiments. It would be interesting to
study the cause for this apparent symmetry breaking.

The selection of which OP is nonzero in the $(C,C)$ phase is a
result of broken symmetry. An electric field along $b$ would order
$P_b$, and then (\ref{PRMO}) would select either $\sigma^{}_1$ or
$\sigma^{}_2$, depending on the sign of the field. Therefore we
suggest cooling the sample into the FE phase in the presence of a
small electric field along $b$. Depending on the sign of the
electric field one should get either the left- or the right-hand
panel of figure \ref{YMOFIG}. This was indeed confirmed experimentally \cite{Rada08}. (A similar experiment was recently
performed in TbMnO$_3$\cite{YY07}).

\section{Critical phenomena}\label{CP}

All the quantitative results presented above were based on the
Landau expansion and on mean field theory. Although these theories
usually give reasonable predictions far away from critical points,
fluctuations must be included in the critical regimes. We start
with NVO and TbMnO$_3$. In these materials, one first goes from
the P phase into the HTI phase, which is represented by a single
complex OP $\sigma_{\rm HTI}$. Since the free energy only involves
$|\sigma_{\rm HTI}|^2$, it does not depend on the phase of this
complex number, and therefore this transition belongs to the
universality class of the XY model, with the critical exponents of
an isotropic ($n=$2)-component spin model. The transition from the
HTI phase into the LTI phase, at $T_<$, is also continuous. A
priori, $\sigma_{\rm LTI}$ is also a complex number, which would
be described by an XY model. However, as we discussed after
(\ref{Delta}), terms like $[(\sigma^{}_{\rm HTI}\sigma_{\rm
LTI}^\ast)^2+{\rm c.c.}]$ would lock the wave vectors of the two
order parameters to each other, even before one reaches $T_<$.
This lock-in is indeed observed experimentally in the LTI phases
of NVO\cite{MK06} and  TbMnO$_3$ \cite{MK05}.

Technically, near $T_<$ we have a finite order parameter $\langle
\sigma^{}_{\rm HTI}\rangle \equiv a e^{i\alpha}$. Writing also
$\sigma^{}_{\rm LTI} \equiv e^{-i\alpha}(b+ic)$, the above locking
 term thus becomes $a^2(b^2-c^2)$. Therefore, the real
order parameters $b$ and $c$ now have different quadratic terms,
and only one of them (depending on the sign of the overall
coefficient) orders at a temperature slightly above the `bare'
$T_<$. As stated above, the fixed length constraint prefers
$\sigma^{}_{\rm HTI}$ and $\sigma^{}_{\rm LTI}$ to have different
phases, which implies that $c$ orders first, and the phases of the
two order parameters differ by $\pi/2$. This  then yields a
helical structure in the LTI phase \cite{TK61,TN67,MK06}.
 Furthermore, this phase
relation is also confirmed by the existence of a ferroelectric
moment in the LTI phase, which would not exist if $\phi_{\rm
HTI}=\phi_{\rm LTI}$ (namely if $b$ were to order, rather than
$c$), see (\ref{METWO}). Thus, the transition from HTI to LTI
belongs to the Ising ($n=1$) universality class. Further away from
the critical point the critical exponents may approach their mean
field values $\gamma=1$ and $\beta=1/2$.

We next consider the ME interaction, (\ref{TWOOP}) and
(\ref{METWO}). Assuming that indeed only $c$ orders, we find that
near the HTI$\rightarrow$LTI transition one can replace
(\ref{METWO}) by
\begin{eqnarray}
V_{\rm int}&=& 2r_b a c P_b.\label{VLTI}
\end{eqnarray}
This immediately implies that the actual order parameter at this
transition is not just $c$, but rather a linear combination of $c$
and $P_b$ \cite{SG70,ABH07a}. This implies that the dielectric
constant should diverge near $T_<$, as $\epsilon_b \sim
|T-T_<|^{-\gamma}$, with the Ising susceptibility exponent
$\gamma$. However, as noted before (\ref{TWOOP}), $\chi_E^{-1}$ is
much larger than $|T-T_<|$, and therefore the amplitude of this
divergent term (related to the amplitude of $P_b$ in the mixed OP)
can be quite small. It would be useful to search for this
divergence experimentally. Similarly, we expect that {\it both}
$c$ {\it and} $P_b$ grow below $T_<$ as $(T_<-T)^\beta$, with the
Ising order parameter exponent $\beta$.

We next turn to RFMO. As discussed in Sec. \ref{RFMO}, the ordered phase
has two complex components of the magnetic OP, $\sigma_1$ and $\sigma_2$,
and therefore altogether we have $n=4$ OP components, as described by
(\ref{FMRFMO}). In fact, this free energy can be written as
\begin{eqnarray}
F&=&(T-T_c)(|\sigma_1(q_z)|^2 +
|\sigma_2(q_z)|^2)+u(|\sigma_1(q_z)|^4 +
|\sigma_2(q_z)|^4)\nonumber\\
&+&{\tilde v}|\sigma_1(q_z)|^2
|\sigma_2(q_z)|^2.
\end{eqnarray}
This can be viewed as the free energy of two XY models (with OP's
$\sigma_1$ and $\sigma_2$), which are coupled by the last term. In
terms of the renormalization group (RG), this model has two
competing fixed points: the isotropic $(n=4)$ one with ${\tilde
v}=2u$, and the decoupled one with ${\tilde v}=0$\cite{AA}. It
turns out that $v$ is slightly relevant near the isotropic fixed
point, and ${\tilde v}$ is slightly irrelevant near the decoupled
fixed point, so that as $T$ approaches $T_<$ one could follow two
scenarios. If $v={\tilde v}-2u<0$, iteration would make it more negative, and
one could end up with a crossover from the isotropic $(n=4)$
critical behavior to the asymptotic behavior of two decoupled XY
models.
 However,  this crossover is very slow. Therefore, one might either observe
 effective exponents close to those of the isotropic $(n=4)$ critical behavior,
 or one might encounter relatively large corrections to the decoupled critical
 behavior, due to the irrelevant parameter ${\tilde v}$,
 which would be renormalized into ${\tilde v}(T_<-T)^{-\alpha}$, where $\alpha$
 is the specific heat exponent of the XY model.
Alternatively, if $v>0$ then $v$ would grow larger under
iterations, and one would never reach the vicinity of the stable
fixed point at ${\tilde v}=2u+v=0$. In this case, one probably
ends up with a slow crossover to a weak first order transition.

The ME interaction in RFMO is given in (\ref{PRMO}). Thus, $P_c
\sim \langle |\sigma_1(q_z)|^2 - |\sigma_2(q_z)|^2 \rangle$. The
RHS of this relation represents an order parameter anisotropy.
Near the isotropic fixed point, this average scales as
\begin{eqnarray}
P_c \sim \langle |\sigma_1(q_z)|^2 - |\sigma_2(q_z)|^2 \rangle
\sim \langle |\sigma_1|^2 \rangle^{\lambda},\label{crit}
\end{eqnarray}
where the exponent $\lambda>1$ is associated with the scaling of
quadratic anisotropy terms near the isotropic $n=4$ fixed point
\cite{AA86}. However, for this result to hold we must have
$\sigma_1\sigma_2=0$, which arises only if $v>0$. As explained above,
in this case we expect a crossover to a weak first order
transition. Thus, as $T$ is increased towards $T_<$ we would
expect a gradual variation from  the mean field result, $P_c \sim
\langle |\sigma_1|^2\rangle$, via the critical behavior of
(\ref{crit}), to a weak first order transition. The mean field
behavior, with $\lambda=1$, implies that the FE moment is
proportional to the intensity of Bragg peaks, as apparently found
experimentally \cite{MK07}. It would be interesting to check this
relation close to $T_<$.

Finally we turn to RMO. As stated, the ordering below the P phase
is into the $(I,I)_1$ phase, which corresponds to a single irrep,
say $\Gamma_a$. As seen from (\ref{F4a}), this ordering involves
the two complex OP's $\sigma_a^+$ and $\sigma_a^-$, and therefore
belongs to some $n=4$ universality class. In the $(I,I)_1$ phase,
where $q_z\ne 1/4$, the quartic terms in the free energy include
only those with the coefficients $c_1$ and $c_2$. Clearly, this
free energy is equivalent to the one discussed above for RFMO,
yielding only one wave vector if $c_2>0$ and two wave vectors if
$c_2<0$. In the former case one probably flows under the RG
towards a weak first order transition, while in the latter case
one would flow towards the stable decoupled fixed point. Thus, the
question whether one or two wave vectors order is directly related
to the nature of the critical behavior.

The situation changes in the $(I,C)_1$ phase, where one also needs
to include the {\it Umklapp} term with $c_3$. Near the decoupled
fixed point, this term involves products of anisotropies in each
of the XY models, and thus it can be shown to be relevant
\cite{AA}. As far as we know, this free energy has no stable fixed
point, and one would eventually end up with a weak first order
transition. However, in the vicinity of the isotropic fixed point,
where $c_2$ and $c_3$ are small, one could still observe the
critical exponents of the isotropic $n=4$ universality class.
In any case, in the generic case the phase $(I,C)_1$ is reached from the phase $(I,I)_1$
via a first order transition, so that the critical behavior of the former can only be
expected near the multicritical point where $J_z=J_{zc}$.

We next discuss the transition into the (so far unobserved)
$(I,I)_2$ phase. We start with the simple case, where only
$\sigma_a^+$ orders in the $(I,I)_1$ phase. As explained after
(\ref{V2EFFEQ}), the locking of the wave vectors of $\sigma_a^+$
and $\sigma_b^+$ fixes the phase of $\sigma_b^+$, so that the
transition into the $(I,I)_2$ phase now involves an Ising-like
order parameter. The situation now becomes exactly the same as in
(\ref{VLTI}): the dielectric constant $\epsilon_b$ would diverge
with the Ising exponent $\gamma$, and $P_b$ would grow in the
$(I,I)_2$ phase with the Ising exponent $\beta$.

The transition from $(I,I)_1$ into $(I,C)_2$ is also weakly first
order, since it involves a lock-in of $q_z$. However, if the
discontinuity is small (as seems to be the case experimentally),
we can still discuss criticality of the OP's associated with
$\Gamma_b$. As discussed after (\ref{V2EFFEQ}), this ordering
should also belong to the Ising universality class: before one
reaches this transition one should see  $\epsilon_b \sim
|T-T_{cb}'|^{-\gamma}$ and $P_b \sim
(T_{cb}'-T)^{\beta}$, with Ising exponents. Since
$\Delta$ now introduces several additional quadratic terms in the
$\sigma_b$'s, this transition is expected to occur at a
temperature $T_{cb}'$ higher than $T_{cb}$, where one would have the
$(I,I)_1\rightarrow (I,I)_2$ continuous transition.

  Near the
P$\rightarrow (I,I)_1$ transition (which occurs at $T_{C1}$), a leading fluctuation expansion
yields $\Delta\epsilon \propto \langle P_b^2\rangle \propto
|\langle \sigma^{2}_a\rangle \langle \sigma^{2}_b\rangle |$.
Since only $\sigma^{}_a$ becomes critical there, we expect
singularities in $\epsilon$ which behave as the energy
($|T-T_{C1}|^{1-\alpha}$) and as the square of the OP
($(T_{C1}-T)^{2\beta}$), but with the appropriate effective $n=4$
exponents. Indeed, experiments\cite{CRC07} show a break in slope
at $T_{C1}$, apparently confirming this prediction. This behavior
is also expected for other multiferroics and indeed this may
explain the anomaly seen in the dielectric constant of NVO shown in
Fig. 4b of \cite{FOCUS}.  In addition,
this anomaly in the zero frequency dielectric constant reflects
the emergence of a resonance in the frequency-dependent dielectric
constant due to electromagnons\cite{AP06,RVA1,RVA2}.

\section{Summary}

 We have developed a phase diagram to
explain the multiferroic behavior of the family of 125's systems
and have proposed several experiments to explore the unusual
symmetries of these systems. In view of our current understanding
it seems unnecessary to invoke the alternate route to
multiferroicity proposed in \cite{JJB07}, particularly as a
microscopic calculation\cite{LCC06} having exactly the symmetry we
have invoked reproduces the experimental data for YMn$_2$O$_5$
quite well.

\ack

We thank  M. Kenzelmann and S. H. Lee for helpful interactions. AA and OEW acknowledge support from the ISF.

\vspace{0.2 in} \noindent
* Also emeritus, Tel Aviv Univesity.

\vspace{1cm} \noindent{\bf References} \vspace{.5cm} 

\end{document}